\documentclass[a4paper,fleqn]{cas-dc}

\usepackage[authoryear]{natbib}

\usepackage{float}
\usepackage{stfloats}
\usepackage{lipsum}
\usepackage{multicol}

\def\tsc#1{\csdef{#1}{\textsc{\lowercase{#1}}\xspace}}
\tsc{WGM}
\tsc{QE}

\begin{document}
\let\WriteBookmarks\relax
\def\floatpagepagefraction{1}
\def\textpagefraction{.001}

\title[mode = title]{Implantation of asteroids from the terrestrial planet region: The effect of the timing of the giant planet instability}


\shorttitle{Asteroid belt implantation}    

\shortauthors{Izidoro et al.}

\author[1]{André Izidoro}[orcid=0000-0003-1878-063]

\cormark[1]


\ead{izidoro@rice.edu}



\affiliation[1]{organization={Department of Earth, Environmental and Planetary Sciences,  Rice University},
            addressline={6100 Main St., MS 126}, 
            city={Houston},
            postcode={77005-1827}, 
            state={Texas},
            country={USA}}

\author[2]{Rogerio Deienno}[orcid=0000-0001-6730-7857]
\affiliation[2]{organization={Southwest Research Institute},
            addressline={ 1050 Walnut St. Suite 300}, 
            city={Boulder},
            postcode={80302}, 
            state={CO},
            country={USA}}

\author[3]{Sean N. Raymond}[orcid=0000-0001-8974-0758]
\affiliation[3]{organization={Laboratoire d'Astrophysique de Bordeaux, Univ. Bordeaux, CNRS},
                addressline={B18N, all{\'e}e Geoffroy Saint-Hilaire},
                city={Pessac},
                postcode={33615},
                country={France}}

\author[4]{Matthew S. Clement}[orcid=0000-0001-8933-6878]
\affiliation[4]{organization={Johns Hopkins APL},
                addressline={11100 Johns Hopkins Rd},
                city={Laurel},
                postcode={20723},
                state={MD},
                country={USA}}                




\begin{abstract}
The dynamical architecture and compositional diversity of the asteroid belt strongly constrain planet formation models. Recent Solar System formation models have shown that the asteroid belt may have been born empty and later filled with objects from the inner ($<$2~au) and outer regions (>5 au) of the solar system. In this work, we focus on the implantation of inner solar system planetesimals into the asteroid belt - envisioned to represent S and/or E- type asteroids - during the late-stage accretion of the terrestrial planets. It is widely accepted that the solar system's giant planets formed in a more compact orbital configuration and evolved to their current dynamical state due to a planetary dynamical instability.  In this work, we explore how the implantation efficiency of asteroids from the terrestrial region correlates with the timing of the giant planet instability, which has proven challenging to constrain. We carried out a suite of numerical simulations of the accretion of terrestrial planets considering different initial distributions of planetesimals in the terrestrial region and dynamical instability times. Our  simulations show that a giant planet  dynamical instability occurring at $t\gtrapprox5$ Myr  --relative to the time of the sun's natal disk dispersal--  is broadly consistent with the current asteroid belt, allowing the total mass carried out by S-complex type asteroids to be implanted into the belt from the terrestrial region. Finally, we conclude that an instability that occurs coincident with the gas disk dispersal is either  inconsistent with the empty asteroid belt scenario, or may require that the gas disk in the inner solar system have dissipated at least a few Myr earlier than the gas in the outer disk (beyond Jupiter's orbit).


 
 




%

\end{abstract}

\maketitle



\section{Introduction} \label{sec:intro}
The main asteroid belt is located between the orbits of Mars and Jupiter -- extending from $\sim$2.1 to $\sim$3.3~au --  and its origin remains heavily debated \cite[e.g.][]{izidororaymond18,raymondnesvorny22}. Although there is some level of radial segregation among the different taxonomic type of asteroids in the belt, the orbital distributions of different groups broadly overlap. The inner part of the main belt is mainly populated by silicate-rich S-complex type asteroids whereas the outer part is mostly populated by carbon-rich C-complex type asteroids \citep{gradietedesco82,demeocarry14}. It is very unlikely that all these asteroid taxonomic types formed at their current locations, since the distinct chemical and isotopic compositions of meteorites associated with these asteroids strongly suggest that they formed in very different environments \citep{warren11,buddeetal2016,Kruijeretal17}. S-complex type asteroids are associated with non-carbonaceous type meteorites. C-type asteroids are associated with carbonaceous-type meteorites \citep[e.g.][]{vernazzapierre16}.  Carbonaceous and non-carbonaceous meteorites show distinct stable isotopic compositions for a large number of non-volatile elements as  siderophile Mo, W, Ni, Ru and lithophile Cr, Ti \citep{warren11,Kruijeretal17}, and the volatile element N \citep{grewaletal21}, suggesting spatial separation at the timing of their formation \citep{kruijeretal20}. This view is consistent with our current understanding of solar system formation and evolution. 

Solar System formation models have showed that asteroids  may have been implanted and trapped  into belt -- from the inner ($<$1-1.5 au) and outer regions ($>$5 au) of the solar system --  via different dynamical  processes. Implantation and trapping mechanisms include the growth and migration of the gas giant planets \citep[e.g.][]{raymondizidoro17a}, gravitational scattering during the accretion of the terrestrial planets \citep{bottkeetal06,raymondizidoro17b}, and the effects of resonances associated with the giant planets \citep{levisonetal09,vokrouhlickyetal16,raymondizidoro17b}. In this work, we focus on the implantation of inner solar system asteroids into the belt -- which are envisioned to represent S-complex type asteroids --  during the so-called late state of accretion of terrestrial planets.   For studies on the implantation of asteroids from the outer solar system (initially beyond Jupiter and Saturn),  we refer the reader to previous studies \citep{walshetal11,raymondizidoro17a,levisonetal09,vokrouhlickyetal16}.  

\subsection{Previous studies}

Studies of asteroid implantation during the late stage of accretion of terrestrial planets come in two flavours. They have been either inspired  by the classical scenario of terrestrial planet formation \citep{chambers01,raymondetal06,obrienetal06}, which assumes wide distributions of planetesimals in the terrestrial region extending from $\sim$0.5 to $\sim$4~au \citep{bottkeetal06}, or instead build on scenarios where terrestrial planets  form from  narrow rings of planetesimals (e.g. $\sim$0.7 to $\sim$1-1.5~au).  A number of studies have shown that the  classical scenario of terrestrial planet formation fails to match important solar system constraints  -- as the low mass of Mars \citep{agnor99,chambers01,hansen09,raymondetal09,obrienetal06} --  and should be abandoned. Therefore, in this work, we focus instead on the formation of terrestrial planets from narrow rings of planetesimals \citep{hansen09,drazkowskaalibert17,izidoroetal22,morbidellietal22}. For a recent review about terrestrial planet formation models we refer the reader to \cite{raymondetal20}.

It is traditionally considered that at the onset of the so-called late stage of accretion of terrestrial planets, the sun's natal gaseous disk has already dissipated and Jupiter and Saturn are already fully formed \citep{chambers01}. It is also widely accepted that the solar system giant planets  formed in a more compact orbital configuration and eventually evolved to their current dynamical state due to a planetary dynamical instability \citep{gomesetal05,levisonetal11,nesvornymorbidelli12}. Yet, the timing of the instability itself remains weakly constrained.

Recent cosmochemical and dynamical arguments have been used to suggest that the solar system dynamical instability happened no later than $\sim$100~Myr after the formation of the solar system \citep{morbidellietal18,nesvornyetal18,ribeiroetal20}, and potentially as early as $\sim$5-10~Myr \citep{ribeiroetal20,liuetal22,nesvornyetal23}. We refer to this scenario ($<$100~Myr) as the ``early instability'' model \citep{clementetal18}. In contrast, in the ``late instability''  model, Jupiter and Saturn are thought remain on compact and almost circular and coplanar orbits during the main phase of accretion of terrestrial planets and beyond ($\gtrapprox$100-500~Myr; \cite{gomesetal05}). A dynamical instability occurring after the terrestrial planets are fully formed (30-150 Myr after the formation of the solar system; \cite[e.g.][]{yinetal02,kleinetal09}) may dynamically over excite or even destabilize the terrestrial planets \citep[e.g.][]{kaibchambers16}.  Yet, to our knowledge, the implantation of asteroids from narrow rings of planetesimals  into the belt has been only comprehensively studied in the context of the ``late instability'' model \citep{raymondizidoro17a}.   In this work, we revisit this problem by testing how the timing of the instability  influences the implantation of asteroids in the belt. 



The dynamical topology and structure of mean motion and secular resonances in the belt are very different when the giant planets have either pre or post-instability (current) orbits \citep[e.g.][]{morbidellihenrard91,izidoroetal16}. This affects how asteroids are scattered from the ring, how they get transported, and finally trapped into the belt.  The primary goals of our current study are to determine whether a sufficient amount of S-complex type asteroids (envisioned to sample the inner solar system) can be implanted in the belt in the early instability scenario ($\lesssim10-100$~Myr), and simultaneously investigate the possible use of implantation as a constraint on the instability's timing. In this work we address these questions by performing a suite of numerical simulations that cover a large range of possible scenarios to comprehensively probe how the timing of the instability correlates with the efficiency of implantation and the final distribution of planetesimals -- envisioned to be asteroids -- in the belt. Our model is not designed to comprehensively reproduce the dynamical architecture of the asteroid belt because we are particularly interested on the  implantation of inner solar system asteroids. In future work, we will combine this study with the implantation of C-type asteroids.

This paper is organized as follows. In section \ref{sec:methods} we describe our methods. In section \ref{sec:results}, we present our main results. Finally, in Section \ref{sec:conclusion} we summarize our main findings.

\section{Methods}\label{sec:methods}


In this work, we carry out numerical simulations of the accretion of the solar system terrestrial planets. Our simulations are designed to qualitatively follow the results of dust coagulation and planetesimal formation models suggesting that the terrestrial planets accreted from a narrow ring of planetesimals located at around 1 au \citep{hansen09,izidoroetal14a,drazkowskaalibert17,izidoroetal16,izidoroetal22,morbidellietal22}.  We do not perform dust coagulation and planetesimal formation simulations here, but simulations start from initial distributions of planetesimals qualitatively consistent with results of these models \citep{izidoroetal22}.

We model the growth of D$\approx$100-km sized planetesimals via collisions in rings radially extending from $\sim$0.7 to $\sim$1.5~au. Our simulations consider different radial mass distributions. We assume rings with surface density profiles proportional to $\Sigma(r)=r^{-x}$, were x is set to as 0, 1, and 5.5. For completeness, we also model a ``upside-down U-shape'' ring profile represented by $\Sigma(r)=(-200(r/{\rm au}-1)^2+24){~\rm g/cm^2}$ \citep{izidoroetal22}. Dust coagulation and planetesimal formation simulations produce rings of planetesimals with a variety of radial slopes \citep{drazkowskaalibert17,izidoroetal22}. The main motivation for testing these  different scenarios come from the results of dust evolution and planetesimal formation simulations \citep{drazkowskaalibert17,izidoroetal22,morbidellietal22}, which shows that the ring's radial mass distribution is highly sensitive to the model parameters.
 In all our simulations, we neglect the effects of pebble accretion \citep[see discussion in][]{izidoroetal21}.

Our simulations are conducted in two phases, which we refer to as ``phase 1'' and ``phase 2''. In phase 1, we use the LIPAD code \citep{levisonetal12} to  model the initial phase of growth, namely from planetesimals to planetary embryos (e.g. Moon to Mars-mass planetary objects). Phase 1 starts with 3000 planetesimals \citep{izidoroetal22} and accounts for the presence of an underlying gaseous disk with surface density profile given by $1700 (r/{\rm AU})^{-1}$g/cm$^2$. We neglect the effects of gas-driven type-I migration \citep{wooetal23} onto protoplanets, but we account for gas drag effects onto planetesimals. Planetesimals feel aerodynamic gas drag as modelled in ~\cite{brasseretal07}.  The disk is dissipated following an exponential decay with e-fold timescale of 2 Myr, and is assumed completely dissipated at 5 Myr \citep[e.g.][]{willianscieza11}. In phase 1, we account for collisional evolution of planetesimals, following the algorithm implemented in LIPAD \citep{levisonetal12}. Simulations were integrated considering a timestep of 4 days and neglecting the presence of fully formed giant planets during the first 5 Myr. This later assumption is not expected to affect the quality of our results. At 5 Myr, we stop our phase 1 simulations and we use their outcome as inputs for the phase 2 -- namely the growth phase from planetary embryos to final planets \citep{wetherilletal78,chambers01}. This second phase is integrated using the regular Mercury integration package \citep{chambers99}.

Phase 2 assumes that the gaseous disk has already fully dissipated and that the giant planets are fully formed. This is a common assumption made in models of the late stage of accretion of terrestrial planets \citep{chambers01,raymondetal04,obrienetal06,lykawkaito13}.  Planetary objects with masses smaller than 0.1 Moon-mass at the end of our phase 1 simulations are entered as ``small'' objects in our initial conditions of Mercury.  These objects do not self-interact. Objects more massive than 0.1 Moon-mass are considered  ``big'' objects and gravitionally interact among themselves and with all other bodies. The number of ``big'' bodies (planetary embryos) at the start of our second phase simulations ranges between about 20 and 30, and  the number of ``small'' bodies (planetesimals) ranges between about 300 and 500, depending on the ring surface density profile.

In our pre-instability orbits,  the giant planets are initially placed into the 3:2 mean motion resonance \citep{massetsnellgrove01,morbidellicrida07} with $a_{\rm J}=5.4$~au, $a_{\rm S}=7.3$~au, $e_{\rm J}=e_{\rm S}=0$, and $i_{\rm J}=0$, and $i_{\rm J}=0.5$~deg.  The orbital configuration of Jupiter and Saturn post-instability orbits (current ones) is given by  $a_{\rm J}=5.25$~au, $a_{\rm S}=9.54$~au, $e_{\rm J}=0.048$, $e_{\rm S}=0.056$, and $i_{\rm J}=0$, and $i_{\rm J}=1.5$~deg.

The timing of the giant planet instability is treated as a free parameter in our simulations. The timing of the instability is not expected to strongly affect the final system of terrestrial planets \citep{clementetal18,nesvornyetal2021}, but it should strongly affect asteroid implantation in the belt. With this motivation, we have performed simulations considering that the instability may have happened at 0, 5, 10, 50 and, 100 Myr, relative to the time of the gas disk dispersal (start of ``phase 2''). We have performed 50 numerical simulations for each combination of ring profile and timing of the giant planet instability. All together, we have performed about $>$1000 numerical simulations modelling the  accretion of terrestrial planets.

In order to model the effect of the dynamical instability on the implantation of planetesimal into the belt  we invoke two different approaches. In our nominal approach, we have modelled the giant planet dynamical instability invoking a well-tested giant planet instability evolution, which has been demonstrated to be consistent with several solar system constraints \citep{deiennoetal18}. In order to precisely reproduce the orbits of the giant planets as in \cite{deiennoetal18}, the orbits of the giant planets were evolved using numerical interpolation. As instability simulations are dynamically chaotic and hard to reproduce, this is a technique commonly used in studies of the solar system instability \citep{roigetal21}. We refer to this instability model as ``Interpolated instability''. In the secondary approach, we mimic the giant planet instability by assuming, for simplicity, that the orbits of the giant planets -- namely Jupiter and Saturn --  instantaneously ``jump'' from pre-instability orbits to the current ones at $t=t_{\rm inst}$. We refer to this instability model as ``Instantaneous instability''. This approach is  imperative because  it allow us to isolate  the relative roles of Jupiter and Saturn in affecting planetesimal implantation \citep{bottkeetal12,deiennoetal16}. 

Previous studies have shown that  different dynamical evolutions of the giant planets may lead to broadly good matches to the solar system dynamical architecture \citep{nesvornymorbidelli12,deiennoetal17,gomesetal18,deiennoetal18}. Yet, each of these different evolutions may have a slightly different impact on the dynamics of asteroids in the belt \citep{mintonmalhotra09,morbidellietal10,izidoroetal16,brasiletal16,deiennoetal18,clementetal21}. With these two  approaches at hands, we hope to be able to draw broadly generic and robust conclusions from this study.

\section{Results}\label{sec:results}

\begin{figure*}
\centering
\includegraphics[scale=0.25]{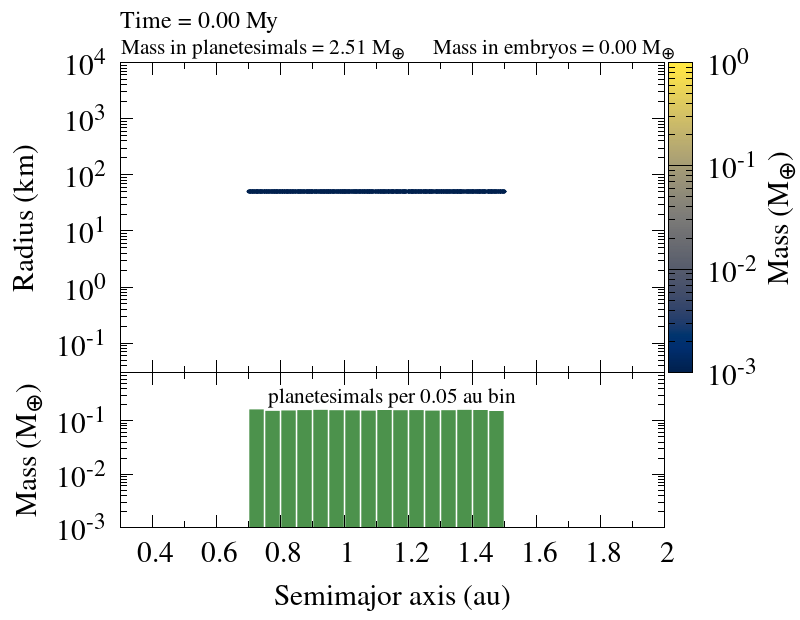}
\includegraphics[scale=0.25]{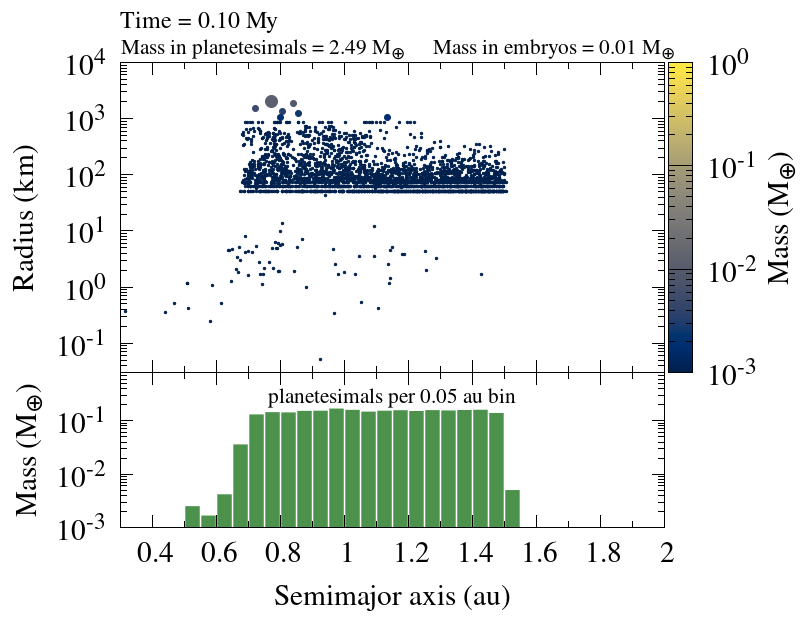}
\includegraphics[scale=0.25]{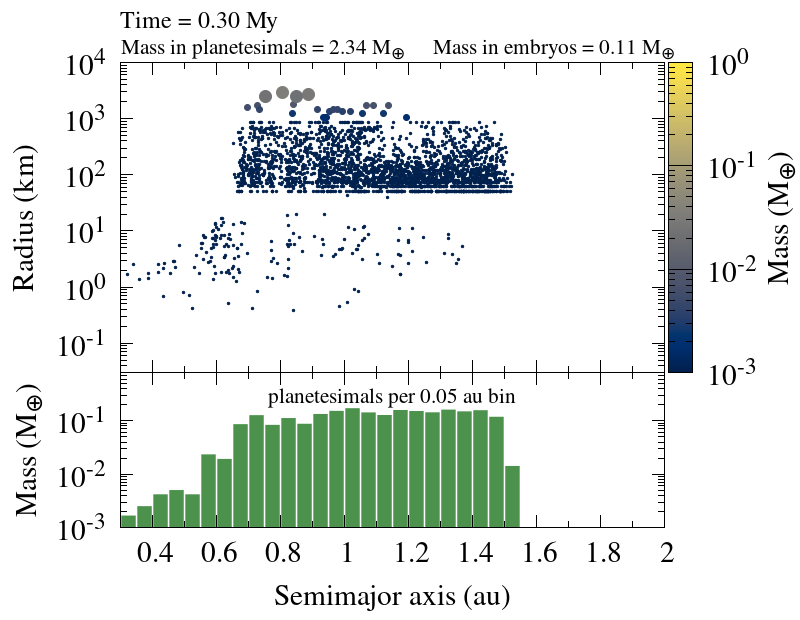}
\includegraphics[scale=0.25]{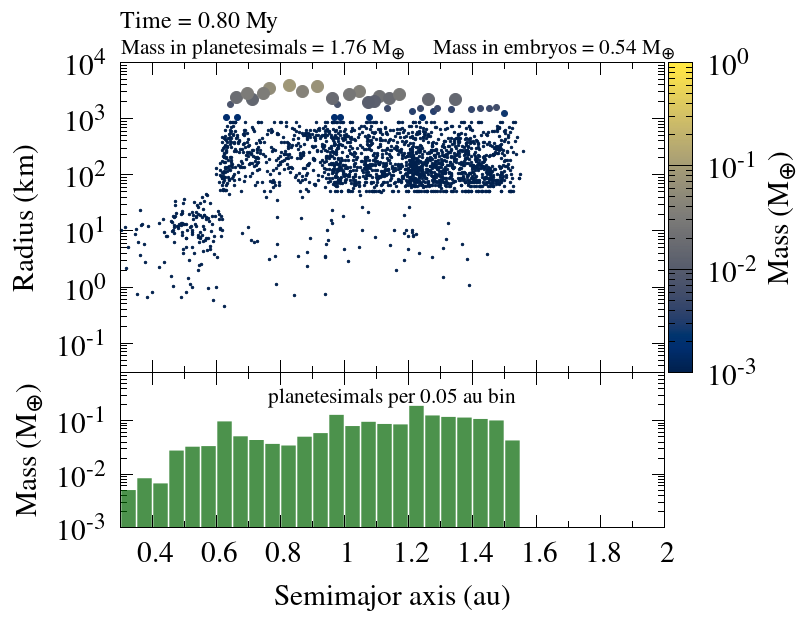}
\includegraphics[scale=0.25]{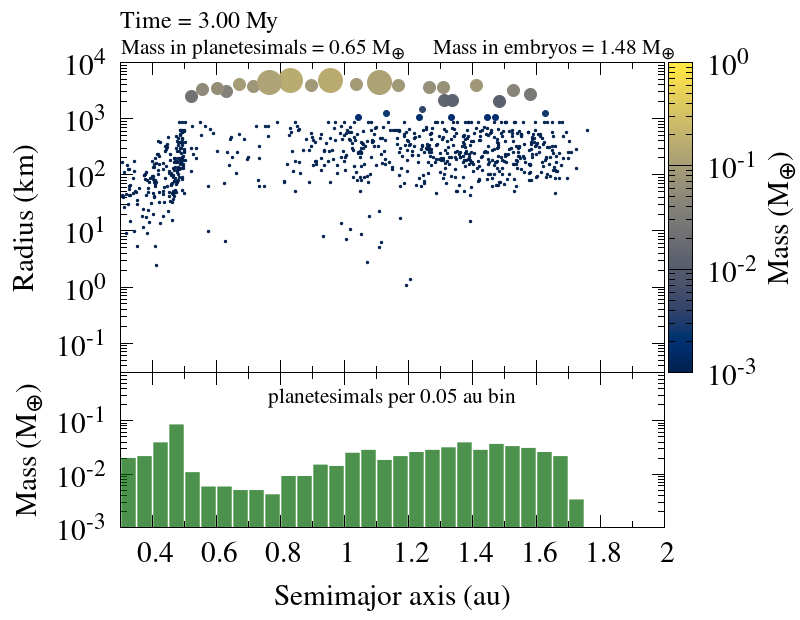}
\includegraphics[scale=0.25]{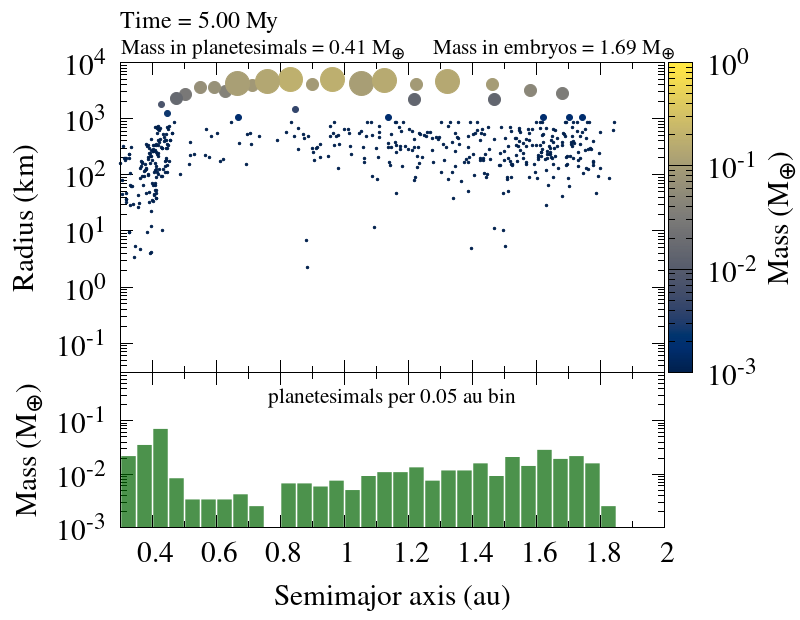}
\caption{Snapshots of the dynamical evolution and growth of planetesimals during the gas disk. The initial ring surface density slope is $x=1$ and the total initial mass carried by planetesimals between 0.7 and 1.5 au is 2.5$M_{\oplus}$. This is a representative example of our phase 1 simulation.  The top-panel of each plot shows the semi-major axis versus objects'  physical radii. The bottom-panel shows the total mass in  planetesimals in different radial-bins. The time of each snapshot is indicated at the top of each plot relative to the start of the simulation. The masses carried by planetesimals and planetary embryos are given at the top of each panel.}
    \label{fig:phase1}
\end{figure*}

Figure \ref{fig:phase1} shows the growth from planetesimals to  planetary embryos in a simulation where the ring surface density slope is $x=1$. This is a representative example of our phase 1 simulations using LIPAD. Planetesimals start with a diameter of 100 km. The disk dissipates at 5 Myr. The final state of phase 1 simulations are used as inputs for our phase 2 simulations.

\begin{figure*}
\centering
\includegraphics[scale=0.28]{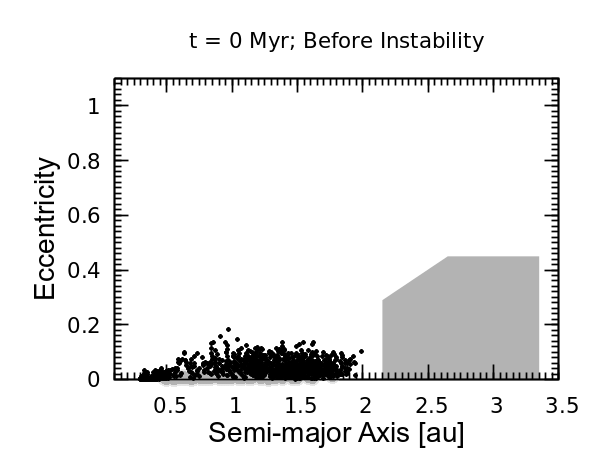}
\includegraphics[scale=0.28]{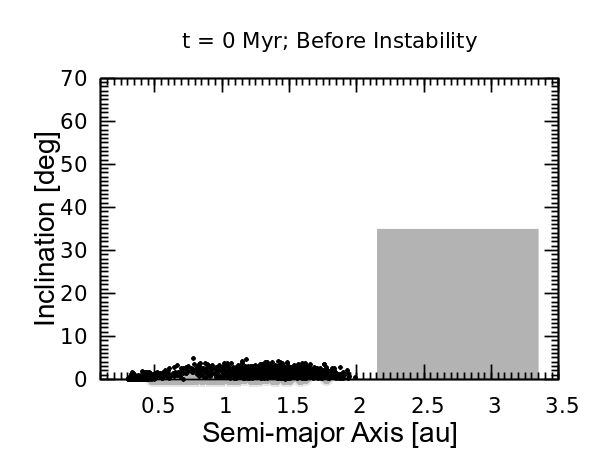}

\includegraphics[scale=0.28]{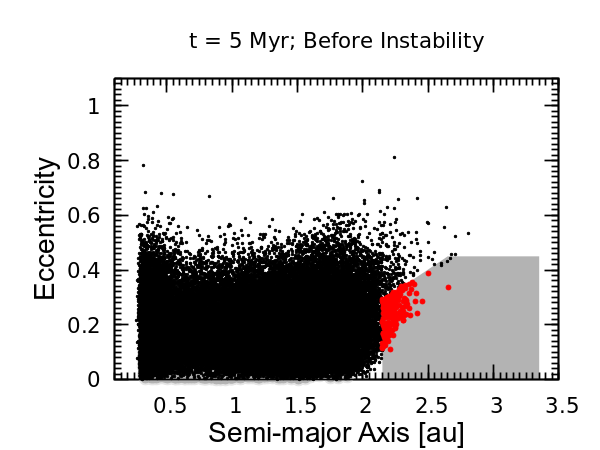}
\includegraphics[scale=0.28]{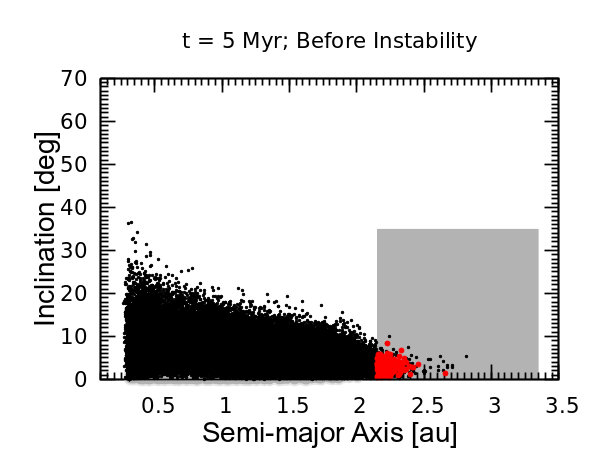}

\includegraphics[scale=0.28]{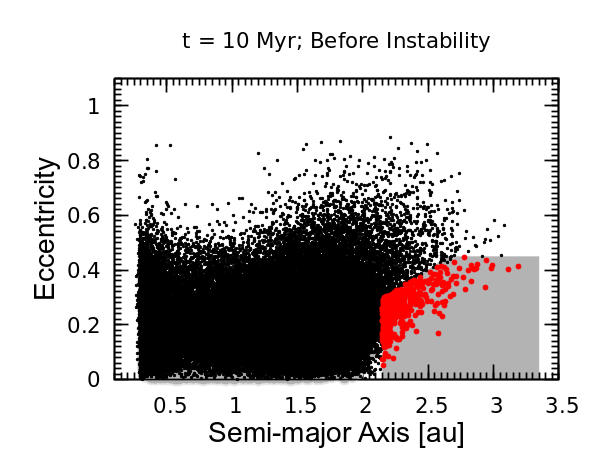}
\includegraphics[scale=0.28]{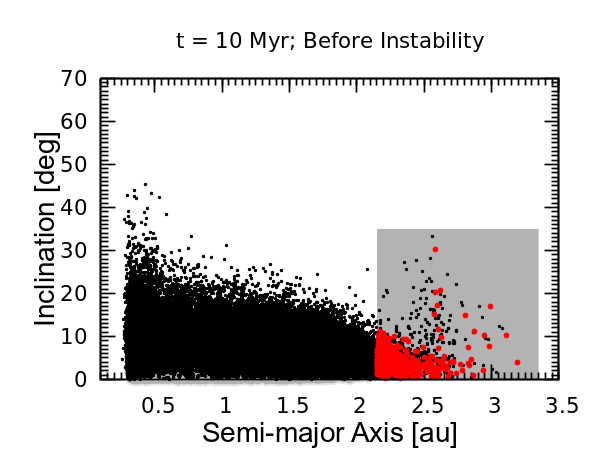}

\includegraphics[scale=0.28]{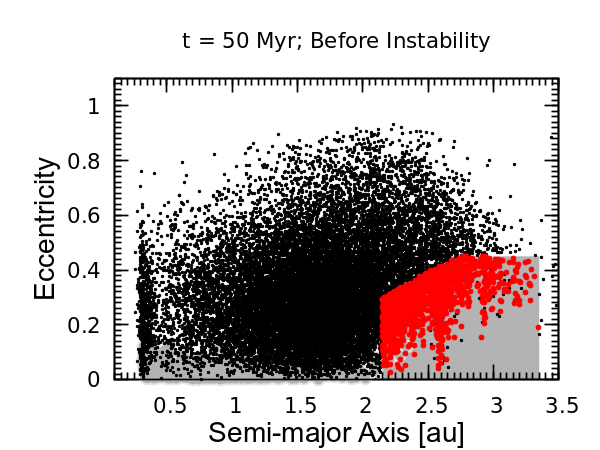}
\includegraphics[scale=0.28]{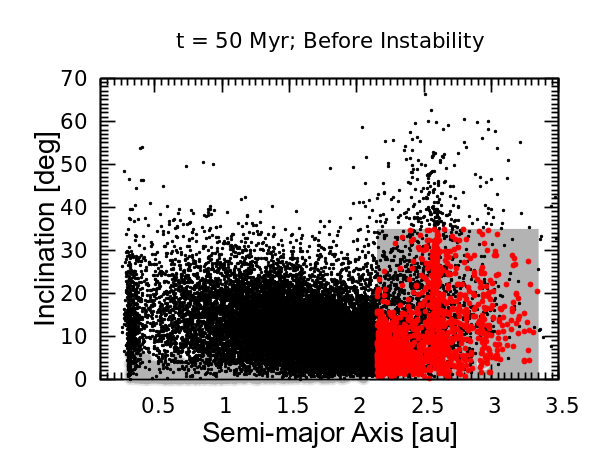}

\includegraphics[scale=0.28]{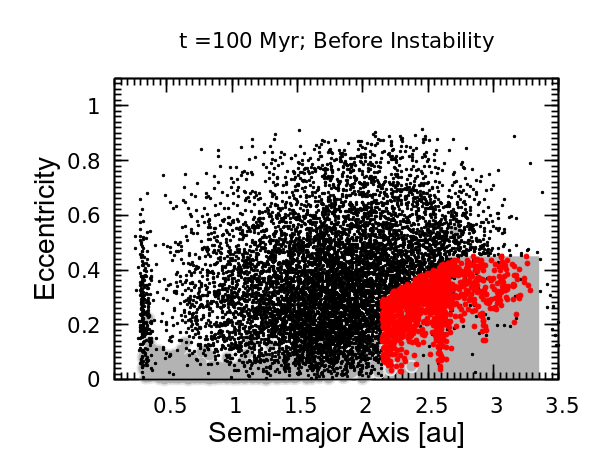}
\includegraphics[scale=0.28]{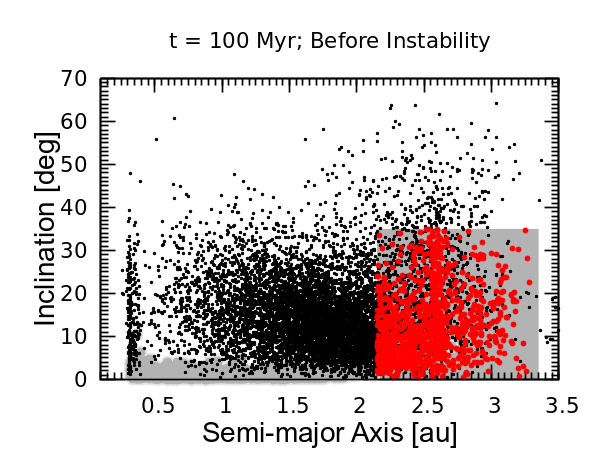}
\caption{Orbital distribution of planetary objects at 0, 5, 10, 50, and 100 Myr in all our simulations, relative to the timing of the gas disk dispersal. Left-panels show  semi-major axis versus eccentricity. Right-panels show semi-major axis versus orbital inclination. Light-grey dots show planetary embryos and planets. Small black dots show  planetesimals. Red dots show planetesimals (``asteroids'') that have been scattered from the ring and implanted into the asteroid belt. In these simulations the planetary instability have not occurred yet and Jupiter and Saturn are in pre-instability orbits.}
    \label{fig:beforeinstability}
\end{figure*}

In order to account for the effects of the giant planet instability, we run our phase-2 simulations from 0 to 100 Myr (relative to the timing of the gas disk dispersal) considering the giant planets on pre-instability orbits (see Methods section for details). Once we have completed these runs, we extract the dynamical state of each of our systems at 0, 5, 10, 50, and 100 Myr creating  a new larger set of initial conditions. These will be used as inputs for our simulations accounting for  the effects of the giant planet instability occurring at different times. Figure \ref{fig:beforeinstability}  shows our collection of initial conditions created using this procedure. Note that the very top panels  of Figure \ref{fig:beforeinstability} correspond the dynamical states of our systems at the end of phase 1, and these are the initial conditions where the instability happens at $t_{\rm inst}=0$.  From top-to-bottom, the second, third, fourth, and fifth row of panels show the initial conditions of our simulations where the instability takes place at $t_{\rm inst}=$ 5, 10, 50, and 100~Myr. We recall that all these times are given relative to the timing of the gas disk dispersal.

The grey areas in Figure \ref{fig:beforeinstability}  delimit the asteroid belt region.  We have defined the asteroid belt as the region where $2.15 \leq  a \leq 3.35$~au, $e<0.45$, $i<35$~deg, $q>1.52$~au, where ``$a$'' represents semi-major axis, ``$e$'' orbital eccentricity, ``$i$'' orbital inclination, and ``$q$'' represent orbital pericenter. We defined the inner, middle, and outer parts of the asteroid belt as the sub-regions where $2.15 \leq  a \leq 2.5$~au, $2.5 \leq  a \leq 3$~au, and $3 \leq  a \leq 3.35$~au, respectively. Planetesimals implanted into the belt are shown as red dots in Figure \ref{fig:beforeinstability}. We remind the reader that the typical number of planetesimals implanted in the belt per simulation is relatively small \citep{raymondizidoro17b}. Each of our simulations start with about 300 to 500 planetesimals and only a few planetesimals get implanted into the belt per simulation, if any. When analyzing our results, we will take advantage from the combined sample of our simulations in order to have a more representative sample \citep{raymondizidoro17b}.

Figure \ref{fig:beforeinstability} shows that at $t=0$ (relative to the gas disk dispersal), planetesimals  and planetary embryos (gray and black dots) have slightly spread out but they remain concentrated inside 2~au. At this point,  no planetesimal has been implanted into the belt in any of our simulations. At $t=5$ and $10$~Myr, the population of planetesimals and planetary embryos have  spread out significantly beyond the original limits of the ring (0.7-1.5~au). The radial spreading of the ring, as planetesimals grow, has been also observed in a number of previous studies \citep{hansen09,walshetal11,deiennoetal19}. At $t=10$~Myr, objects have been scattered inside the asteroid belt region, with some objects reaching even the outer parts of the belt (a$>$3~au). In these simulations, the giant planet instability will take place when planetesimals have already entered into belt. The fate of these planetesimals will be explored in the next section when we will include the effects of the instability on these objects.  At $t=50$ and 100 Myr, the number of planetesimals implanted into the belt is relatively large.

Table \ref{tab:tablebefore} reports  planetesimal implantation efficiencies  just before the instability for different disk slopes and instability times. Efficiencies are calculated by counting the number of red dots in Figure \ref{fig:beforeinstability} and dividing by the initial number of planetesimals in the entire ring. When analyzing the implantation efficiency pre-instability our goal is to understand the effect of the instability on the implantation process. One of the questions we want to answer is: Does the instability foster or hinder planetesimal implantation? Table entries reported with a leading sign ``$<$'' implies that no planetesimal have been implanted and, therefore, the implantation efficiency is per-definition smaller than 1/N, where N represents the total number of planetesimal at the end of the gas disk phase. This is the case, for instance, for all simulations of Table \ref{tab:tablebefore} where $t_{\rm inst}=0$ (see four top  lines). This table also shows that rings with very steep planetesimal surface density profiles (e.g. $x=5.5$, Table \ref{tab:tablebefore}, line 7 from the top) tend to fail in implanting planetesimals in the belt during the first 10 Myr. In this particular scenario, at 10 Myr  growing  planets in the outer parts of the ring ($>$1~au) are too small to significantly scatter planetesimals into the belt.

We now use the systems from Figure \ref{fig:beforeinstability} as inputs for simulations accounting for the effects of the solar system dynamical instability. We will first model the giant planet instability following the interpolation method \citep[e.g.][]{deiennoetal18}, and later in this section we will present the results of ``instantaneous''  instability simulations. We restate that Figure \ref{fig:beforeinstability} shows the results of a collection of planetary systems. We will  perform one independent instability simulation for each of our planetary systems.


\begin{table}
\vspace{1cm}
\centering
\scriptsize
\caption{Implantation efficiency of planetesimals just before the instability in simulations  with different ring surface density profiles and instability times. The columns are ring slope, instability time (Myr), and implantation efficiency in the inner, middle, and outer belt. The last column gives the total implantation efficiency. Each table-row gives the results of 50 numerical simulations over which efficiencies are computed from\label{tab:tablebefore}.}
\begin{tabular}{c c c c c c}
\hline
\multicolumn{1}{c}{}   &   \multicolumn{5}{c}{Implantation efficiency just before the instability}           \\
\multicolumn{1}{c}{Ring}  & \multicolumn{1}{c}{Inst.} &    \multicolumn{3}{c}{}         \\
   slope                   &       time (Myr)                       &                     \multicolumn{1}{c}{Inner}  & \multicolumn{1}{c}{Midlle}  & \multicolumn{1}{c}{Outer}& \multicolumn{1}{c}{Total}\\
      \hline
0         &   0    &     $<$3.93e-5   &     $<$3.93e-5   &     $<$3.93e-5        &   $<$3.93e-5       \\    \hline
1         &   0    &     $<$4.36e-5   &     $<$4.36e-5   &     $<$4.36e-5        &   $<$4.36e-5     \\    \hline
5.5       &   0    &     $<$7.25e-5   &     $<$7.25e-5   &     $<$7.25e-5        &   $<$7.25e-5      \\    \hline
U    &   0    &     $<$5.59e-5   &      $<$5.59e-5  &     $<$5.59e-5        &   $<$5.59e-5      \\    \hline\hline
0         &   5    &    5.85e-03   &      7.85e-05   &     $<$3.93e-5        &    5.93e-03       \\    \hline
1         &   5    &     6.10e-4   &     $<$4.36e-5   &     $<$4.36e-5        &   6.10e-4     \\    \hline
5.5       &   5    &     $<$7.25e-5   &     $<$7.25e-5   &     $<$7.25e-5        &   $<$7.25e-5      \\    \hline
U    &   5    &     $<$5.59e-5   &      $<$5.59e-5  &     $<$5.59e-5        &   $<$5.59e-5      \\    \hline\hline
0         &   10    &     1.30e-2     &   2.00e-3     &     3.93e-5           &  1.51e-2       \\    \hline
1         &   10    &     3.57e-3   &     4.36e-4   &     $<$4.36e-5        & 4.05e-3       \\    \hline
5.5       &   10    &  $<$7.25e-5   &  $<$7.25e-5   &     $<$7.25e-5        &   $<$7.25e-5      \\    \hline      
U    &   10    &     5.59e-4   &     1.68e-4   &     5.59e-5        &   7.82e-4      \\  \hline\hline
0         &   50    &     1.52e-2   &     1.27e-2   &     8.64e-4       &  2.87e-2      \\    \hline
1         &   50    &     9.19e-3   &     7.23e-3   &     1.30e-3    &  1.77e-2       \\    \hline
5.5       &   50    &     2.90e-4   &     1.01e-3   &     7.25e-5       &  1.38e-3      \\    \hline     
U    &   50    &     2.63e-3   &     3.40e-3   &     6.70e-4       &  6.70e-3      \\    \hline \hline

0         &   100    &       1.14e-2   &     1.06e-2      & 1.18e-3 & 2.31e-2    \\    \hline

1         &   100    &       7.54e-3   &     7.19e-3      & 1.00e-3 & 1.57e-2   \\    \hline

5.5         &   100    &       3.62e-4   &    1.16e-3      & 2.90e-4 & 1.81e-3   \\    \hline
U         &   100    &       2.24e-3   &    2.96e-3      & 4.47e-4 & 5.64e-3   \\    \hline
\end{tabular}

\end{table}

\subsection{Interpolated Instability} 

\begin{figure*}
\centering
\includegraphics[scale=0.28]{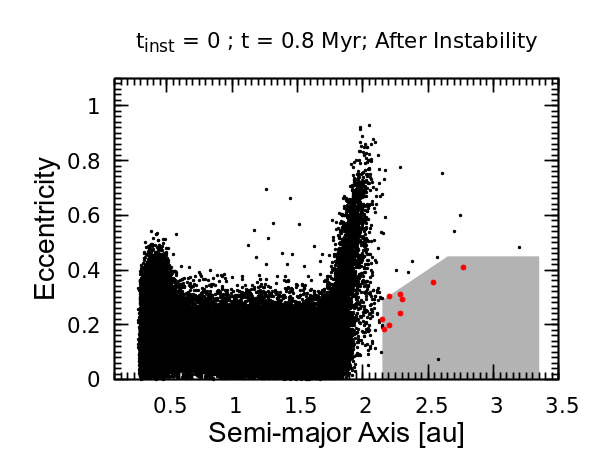}
\includegraphics[scale=0.28]{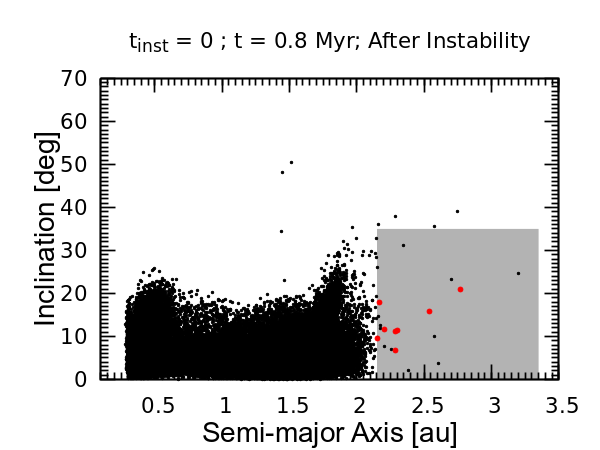}

\includegraphics[scale=0.28]{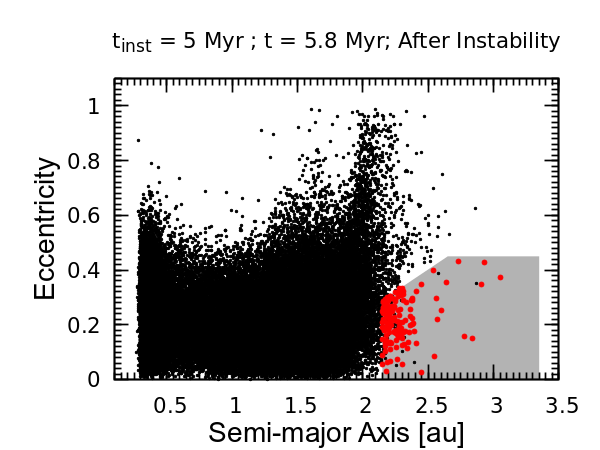}
\includegraphics[scale=0.28]{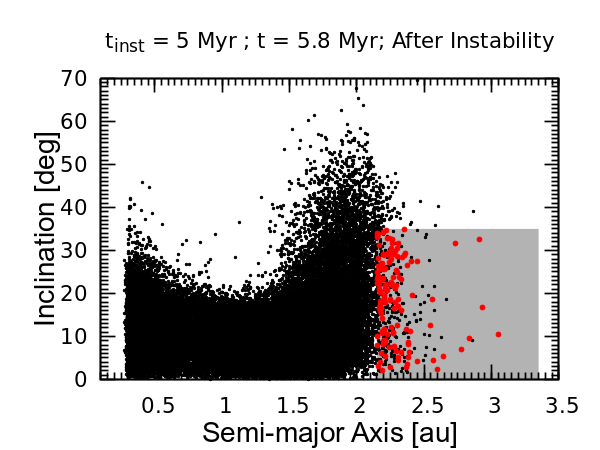}

\includegraphics[scale=0.28]{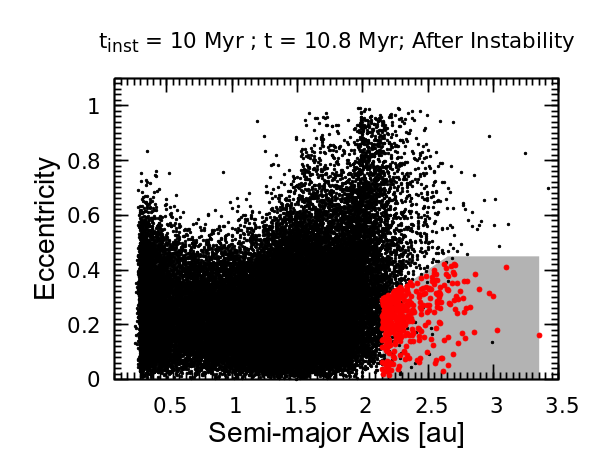}
\includegraphics[scale=0.28]{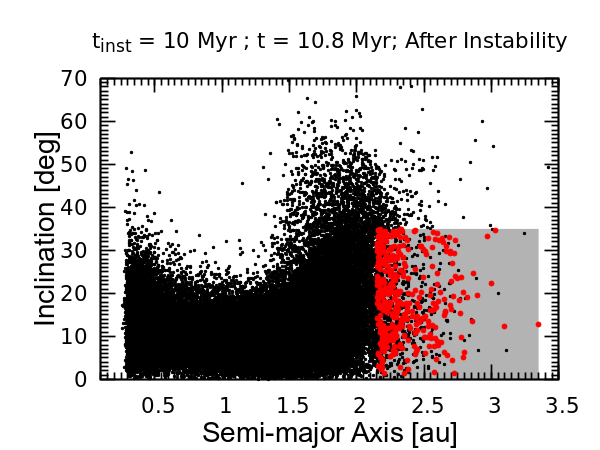}

\includegraphics[scale=0.28]{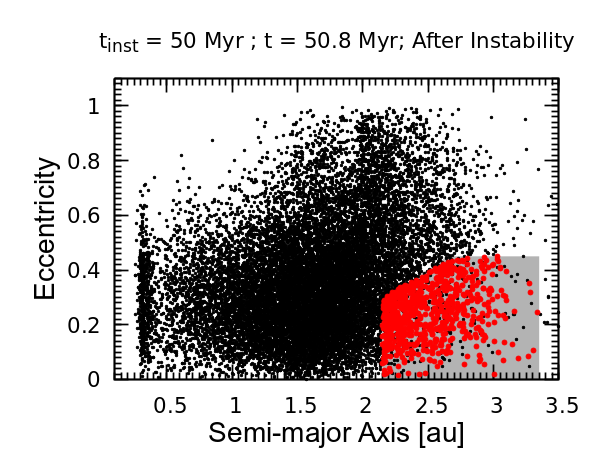}
\includegraphics[scale=0.28]{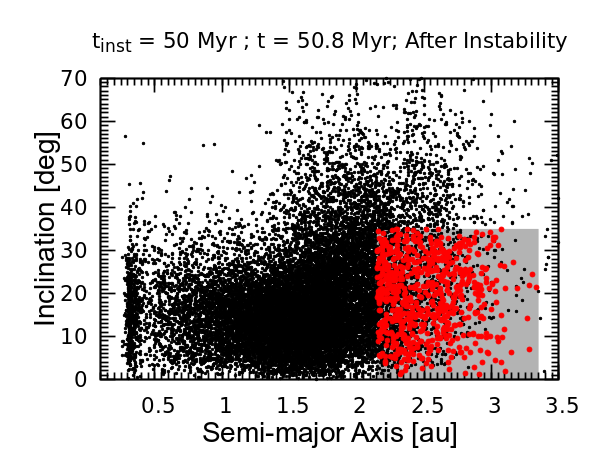}

\includegraphics[scale=0.28]{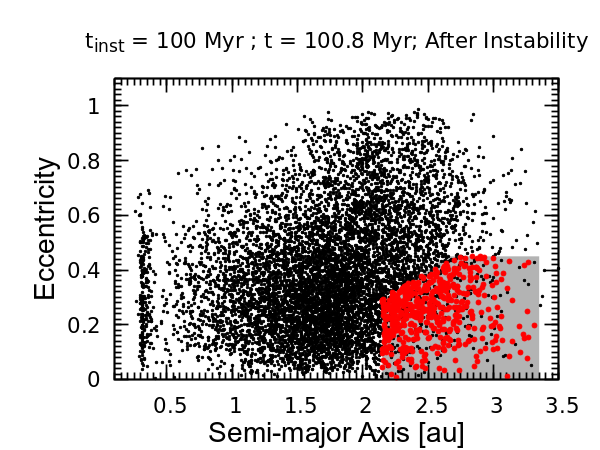}
\includegraphics[scale=0.28]{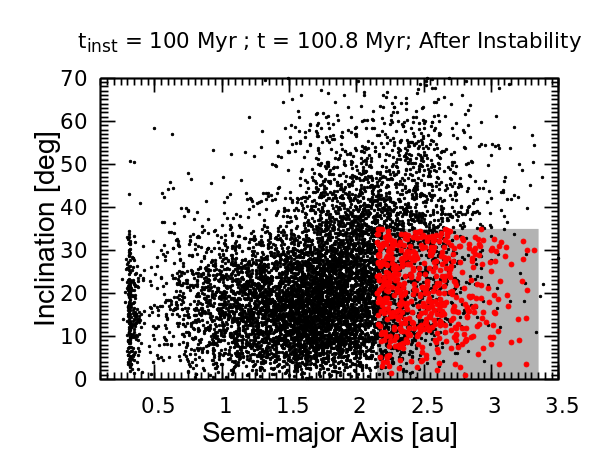}
\caption{Orbital distribution of planetary objects just after the dynamical instability, at t= 0.8, 5.8, 10.8, 50.8, and 100.8 Myr. Each panel-row shows the results of simulations with different instability times. In all cases the instability is modeled using the interpolation approach. Left-panels show  semi-major axis versus eccentricity. Right-panels show semi-major axis versus orbital inclination. Light-grey dots show planetary embryos and planets. Small black dots show  planetesimals. Red dots show planetesimals (``asteroids'') that have been scattered from the ring and implanted into the asteroid belt.}
    \label{fig:after_instability}
\end{figure*}

Figure \ref{fig:after_instability} shows the orbital distribution of the planetary objects of Figure \ref{fig:beforeinstability} just after the giant planet dynamical instability (at the instant that Jupiter and Saturn reach their final orbits). From start-to-end, the instability evolution used in this work lasts about 800~kyr \citep{deiennoetal18}. In other words, Figure \ref{fig:after_instability} shows the orbital configuration of planetary objects at t=$t_{\rm inst}+800$~kyr. 


Comparing Figures \ref{fig:beforeinstability} and \ref{fig:after_instability}, one can see that the instability tend to foster planetesimal implantation in the inner parts of the belt, in particular, in simulations where rings have $x=0$ and $x=1$. Figure \ref{fig:after_instability} shows that one of the roles of the instability is to reshuffle  the orbital distribution of orbits beyond 2.1~au, due to the effects of the chaotic evolution of Jupiter and Saturn, which causes secular variations in planetesimals orbits \citep{izidoroetal16,deiennoetal18}.  During the instability phase, planetesimals may either get implanted or removed from the belt. The implantation efficiencies reported in Table \ref{tab:tableafter}, when compared to those of Table \ref{tab:tablebefore} show this result. Overall, we found that the instability itself may  increase/reduce the implantation efficiency by up to a factor of a few.

We remind the reader that our reported planetesimal implantation efficiencies are computed relative to the number of leftover planetesimal at the time of the gas disk dispersal. Our simulations start with  3000 planetesimals and a total mass of about $\sim2.5M_{\oplus}$, but at the time of the disk dispersal the number of planetesimals  typically have reduced to 300-500 objects (e.g. Figure \ref{fig:phase1}). Consequently, the total  mass carried by planetesimals in our simulations at the end of the gas disk dispersal (leftover planetesimals) is typically about $\sim0.4M_{\oplus}$.

Combining this typical leftover  mass with the implantation efficiency of Table \ref{tab:tableafter} to infer how much mass is ultimately implanted into the asteroid belt in each of our different models is  a tempting approach, but not really correct. One needs to  keep in mind that the total mass carried by planetesimals at the end of the gas disk phase is  model dependent. It depends on the gas disk lifetime and planetesimal initial sizes in our phase 1 simulations, which are not strongly constrained. If we had assumed a longer disk lifetime (or a smaller initial planetesimals size), the total mass in leftover planetesimals at the end of the gas disk phase would be probably smaller \citep{kokuboida00}. On the other hand, disks with shorter lifetimes than 5 Myr (or models assuming larger initial planetesimals sizes) would probably result in a higher total mass leftover in planetesimals \citep{kokuboida00}. We will be able to make more robust conclusions and potentially rule out specific instability timings that implant too little or too much mass in the belt by considering a plausible range of total mass in leftover planetesimals. 

Our phase 1 simulations always start  with 2.5$M_{\oplus}$ in planetesimals in total. The total mass carried by the real terrestrial planets is about 2$M_{\oplus}$. Consequently, in reality, only 0.5$M_{\oplus}$ is truly available to be potentially implanted in the belt because the remaining part has to be used to build the terrestrial planets. This is a firm upper limit. We also know that after the terrestrial planets are almost fully formed about $\sim$0.02-0.05$M_{\oplus}$ in  leftover planetesimals is required to explain the late veneer \citep{raymondetal13} and  lunar crater record \citep{nesvornyetal23}. Consequently, we assume a lower limit  for the total mass available in planetesimals to be implanted in the asteroid belt of 0.05$M_{\oplus}$.

The total mass in the  asteroid belt today is roughly $5\times10^{-4}M_{\oplus}$. About 21\% of this mass ($1.05\times10^{-4}M_{\oplus}$) is carried by S-complex type asteroids \citep{mothe-dinizetal03,demeocarry14}. In order to implant the {\it total} mass carried by S-complex type asteroids it is required --  considering the nominal leftover planetesimal mas in our simulations -- a total implantation efficiency of  $\sim2.625\times10^{-4}$ ($2.625\times10^{-4}\times 0.4M_{\oplus}=1.05\times10^{-4}M_{\oplus})$. Considering instead our envisioned lower and upper limits for the total mass available in planetesimals for implantation, the required implantation efficiency may vary by one order of magnitude, between $\sim2\times10^{-4}$  and  $\sim2\times10^{-3}$. A comparison of these values with those in Table  \ref{tab:tableafter}  can not be made yet and required another important consideration.


In Table \ref{tab:tableafter}, the reported implantation efficiencies are given at the end of the instability phase. Therefore, we should also account for  the associated asteroid belt depletion during the late stage accretion of terrestrial planets (first 200 Myr) and the subsequent solar system evolution after the giant planet instability, over $\sim$4.5 Gyr.

Our results suggest that between the instability and the first 200 Myr of the solar system evolution, the asteroid belt may be depleted by up to a factor of $\sim$5-10 (this will become clear later, when the reader compare Tables \ref{tab:tableafter} and \ref{tab:200Myr}). In addition, Solar System dynamical evolution models suggest that the asteroid belt is depleted by about a factor of $\sim$2, from  $\sim$200 Myr to $\sim$4.5~Gyr \citep{mintonmalhotra09,morbidellietal09b,deiennoetal18}. Assuming a depletion  of a factor of 10 for these two epochs combined, the implantation efficiency required to match the total mass in S-complex type asteroids in Table \ref{tab:tableafter} turns out to be between $\sim2\times10^{-3}$  and  $\sim2\times10^{-2}$. We can now compare these efficiencies with those in Table \ref{tab:tableafter} to rule out specific scenarios (instability times and ring slopes).

Note, however, that if some planetesimals akin to S-complex type asteroids formed in the belt or were implanted via a different mechanism than that considered here~\cite[e.g.][]{raymondizidoro17a}, it would lower our required implantation efficiency.  Planetesimal formation models do allow the formation of planetesimals in the belt when the effects of zonal flows \citep[e.g.][]{gerbigetal19} in the gaseous disk are accounted for \citep{izidoroetal21}.  The efficiency of this mechanism for promoting planetesimal formation, however, remains poorly understood.

In this work, we envision that the asteroid belt was born empty \citep{raymondizidoro17a,izidoroetal22}, therefore, for consistency, we assume that our model should be able to implant the entire population of S-complex type asteroids in the belt (but not too much mass). In Table \ref{tab:tableafter}, simulations where the total implantation efficiency is between $2\times10^{-3}$ and $2\times10^{-2}$, matching our implantation constraint as discussed before, are shown in bold. 

In Table \ref{tab:tableafter}, none of our ring scenarios match this implantation constrain for a dynamical instability occurring at the time of the disk dispersal. For an instability occurring at $t_{\rm inst}=5$~Myr, only our simulations with $x=0$ would be able to implant enough mass into the belt to broadly explain  S-complex type asteroids. For an instability at $t_{\rm inst}=10$~Myr, disks with $x=0$, 1 and upside-down U-shape disk profiles match this constrain. For $t_{\rm inst}\geq50$~Myr, all ring scenarios can implant enough mass into the belt to explain observed S-complex type asteroids.

\begin{table}
\vspace{1cm}
\centering
\scriptsize
\caption{Implantation efficiency of planetesimals just after the instability (``interpolated'' case) in simulations with different ring surface density profiles and instability times. The columns are ring slope,  time (Myr), and implantation efficiency in the inner, middle, and outer belt. The last column gives the total implantation efficiency. Bold values show implantation efficiencies consistent with the total mass carried by S-complex type asteroids in the belt. Each table-row gives the results of 50 numerical simulations over which efficiencies are computed from\label{tab:tableafter}.}
\begin{tabular}{c c c c c c}
\hline
\multicolumn{1}{c}{}   &   \multicolumn{5}{c}{Implantation efficiency just after the instability}           \\
\multicolumn{1}{c}{Ring}  & \multicolumn{1}{c}{Time after} &    \multicolumn{3}{c}{}         \\
   slope                   &      inst. (Myr)                       &                     \multicolumn{1}{c}{Inner}  & \multicolumn{1}{c}{Midlle}  & \multicolumn{1}{c}{Outer}& \multicolumn{1}{c}{Total}\\
      \hline
0         &   0.8    &  2.36e-4     &    7.86e-5   &     $<$3.93e-5        &   3.14e-04        \\    \hline
1         &   0.8    &  4.36e-5     &     $<$4.36e-5   &     $<$4.36e-5      &   4.36e-05        \\    \hline
5.5       &   0.8    &     $<$7.25e-5   &     $<$7.25e-5   &     $<$7.25e-5        &   $<$7.25e-5      \\    \hline
U    &   0.8    &     $<$5.59e-5   &      $<$5.59e-5  &     $<$5.59e-5        &   $<$5.59e-5      \\    
\hline\hline
0         &   5.8    &  4.44e-03     &    3.14e-04   &      3.93e-05        &    {\bf 4.79e-03}       \\    \hline
1         &   5.8    &  6.97e-4     &     4.36e-5   &     $<$4.36e-5      &   7.40e-05        \\    \hline
5.5       &   5.8    &  7.25e-5   &      $<$7.25e-5   &     $<$7.25e-5        &  7.25e-5      \\    \hline
U    &   5.8    &      1.11e-04   &      1.11e-04  &     $<$5.59e-5        &   2.23e-04      \\    
\hline\hline
0           &   10.8     & 8.25e-3     &    2.04e-3    &     3.93e-5        &   {\bf1.03e-2} \\ \hline
1           &   10.8     & 3.36e-3     &    8.27e-4    &     $<$4.36e-5        &  {\bf 4.18e-3}  \\ \hline
5.5       &   10.8    &     7.25e-5   &     $<$7.25e-5   &     $<$7.25e-5        &  7.25e-5      \\    \hline
U         &   10.8    &  1.62e-3     &     4.47e-4   &     1.12e-4     &   {\bf 2.18e-3}       \\    \hline \hline
0           &   50.8     & 7.19e-3     &    6.29e-3    &    5.89e-4        &   {\bf 1.41e-2}  \\ \hline
1           &   50.8     & 5.49e-3     &    4.18e-3    &    3.49e-4        &    {\bf 1.00e-2} \\ \hline
5.5           &   50.8     & 1.09e-3     &    1.16e-3    &    7.25e-5        &   {\bf 2.32e-3}  \\ \hline
U           &   50.8     & 2.63e-3     &    1.96e-3    &    3.35e-4        &   {\bf 4.92e-3} \\ \hline \hline
0           &   100.8     & 5.78e-3     &    4.56e-3    &    5.89e-4        &   {\bf1.09e-2} \\ \hline
1           &   100.8     & 4.01e-3     &   2.96e-3    &    6.10e-4        &    {\bf 7.58e-3} \\ \hline
5.5         &   100.8     & 9.42e-4    &   1.01e-3    &   7.25e-5        &   {\bf 2.02e-3}  \\ \hline
U         &   100.8     & 1.62e-3    &  1.51e-3    &   1.68e-4       &   {\bf 3.30e-3}  \\ \hline

\end{tabular}
\end{table}

We have at this point probed the effect of the dynamical evolution of Jupiter and Saturn during the instability on the implantation of planetesimals into the belt. Our goal now is to probe how the subsequent  accretion of terrestrial planets impacts the implantation efficiency. The natural approach here would be to extend the simulations of Figure \ref{fig:after_instability} up to $\sim$200 Myr timescales, but this is not what we do. Because in this work we test two  giant planet instability scenarios, we extend to 200~Myr only our simulations invoking the ``Instanenous'' dynamical instability. There is not specific motivation for this choice, and our main goal here is to reduce the computer time necessary to run our simulations. Extending both the ``Interpolated'' and ``Instatenous'' simulations batches up to 200 Myr would make our study twice as costly.



\subsection{Instantaneous Instability}

\begin{figure*}
\centering
\includegraphics[scale=0.4]{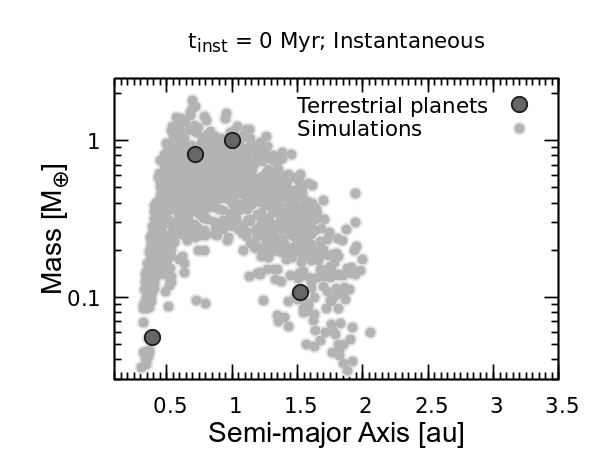}
\includegraphics[scale=0.4]{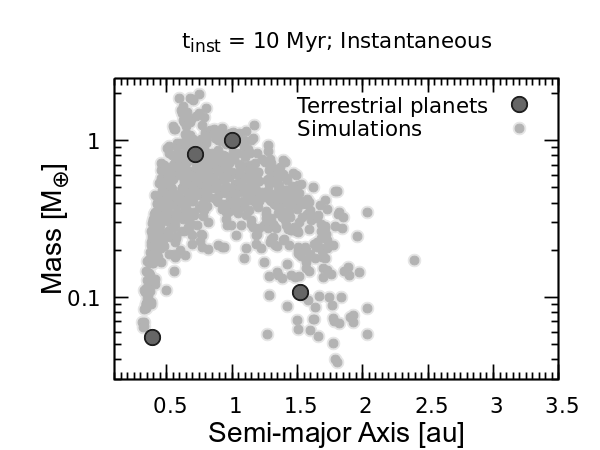}
\includegraphics[scale=0.4]{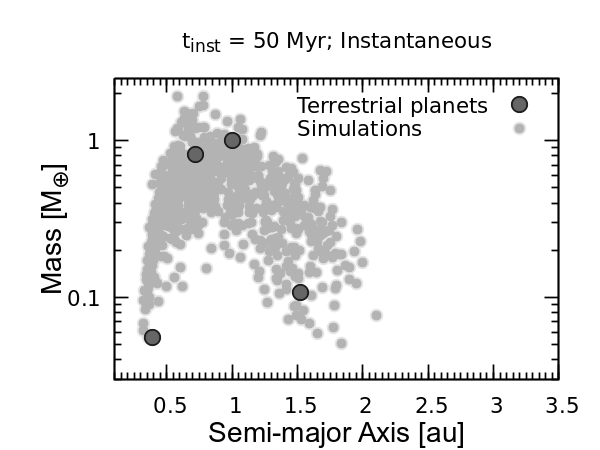}
\includegraphics[scale=0.4]{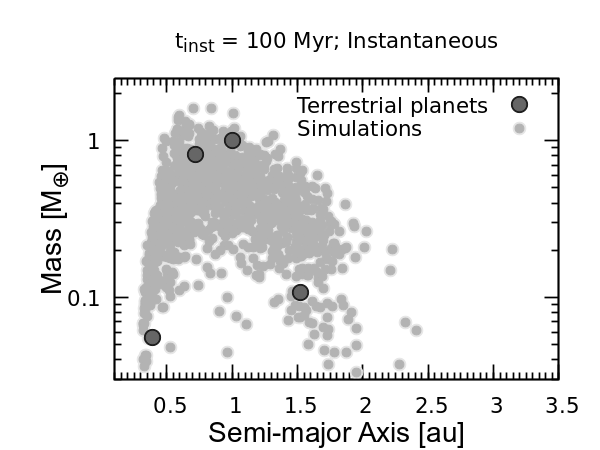}
\caption{Final distributions of planets  at $t=$~200 Myr, in four sets of simulations with different instability times. The timing of the instability is indicated at the top of each panel, relative to the time of the gas disk dispersal. Each panel shows the results of simulations considering rings with different radial surface density profiles. The x-axis shows planets's semi-major axis.  The y-axis shows the planets' mass. Light-grey dots show planets produced in our simulations. The four dark-grey dots represent the real terrestrial planets. The giant planet dynamical instability is mimicked by assuming that Jupiter and Saturn instantaneously jump from their pre-instability orbits to their current ones at $t=t_{\rm inst}$.}
    \label{fig:finalplanets}
\end{figure*}

In this section, we present the results of our simulations  where the instability is modeled assuming that Jupiter and Saturn instantaneously ``jump'' from pre-instability orbits to their current ones. We use the planetary systems from Figure \ref{fig:beforeinstability} as inputs for these simulations.  

Figure \ref{fig:finalplanets} shows the final distribution of planets produced in our  simulations after 200 Myr of integration. As in the previous analysis, we generate these plots by combining the outcome of all ring profiles but  grouping simulations according to their respective timing of the instability ($t_{\rm inst}$, see top of each panel). Each panel shows the results of $\sim$200 simulations. Figure \ref{fig:finalplanets} shows that our simulations match broadly well the masses of all terrestrial planets, producing planets  with masses within a factor of a few of those of the real terrestrial planets, consistent with previous studies \citep[e.g.][]{raymondizidoro17b,clementetal2019}.



\begin{figure*}
\centering
\includegraphics[scale=0.32]{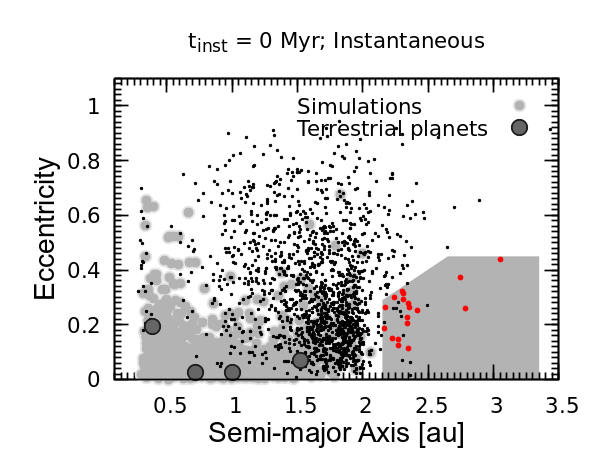}
\includegraphics[scale=0.32]{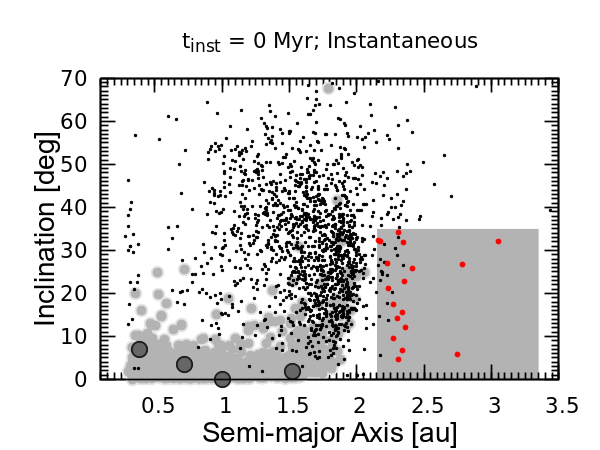}
\includegraphics[scale=0.32]{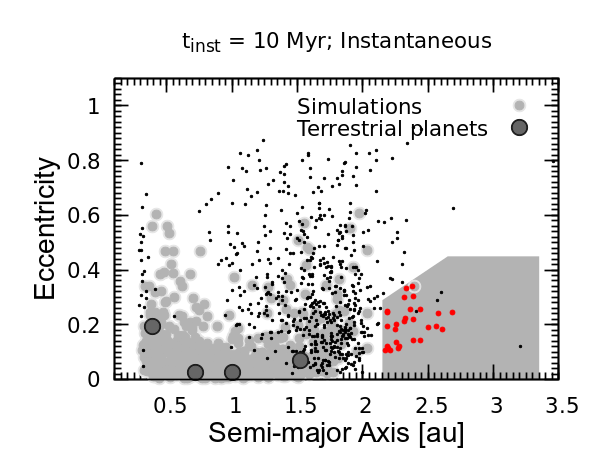}
\includegraphics[scale=0.32]{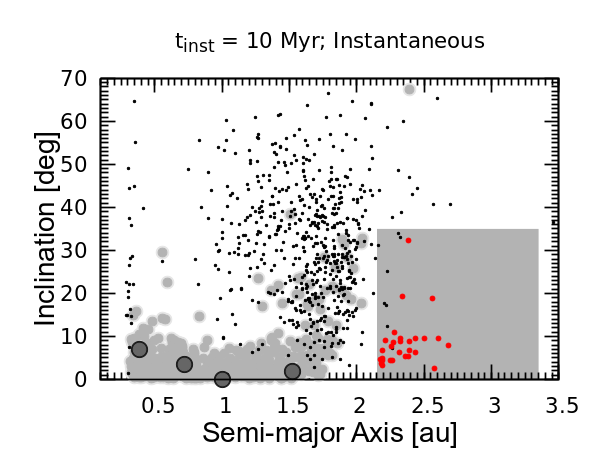}
\includegraphics[scale=0.32]{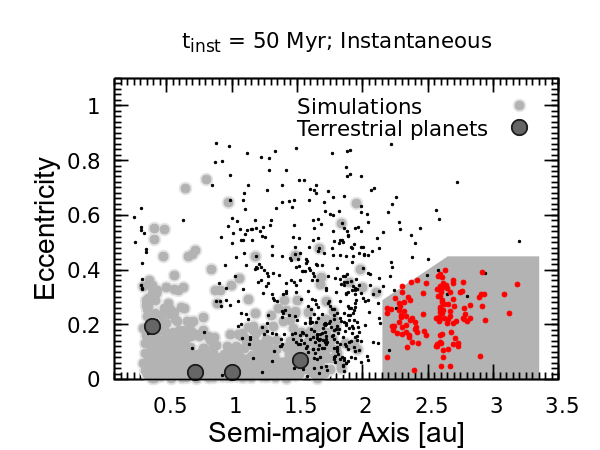}
\includegraphics[scale=0.32]{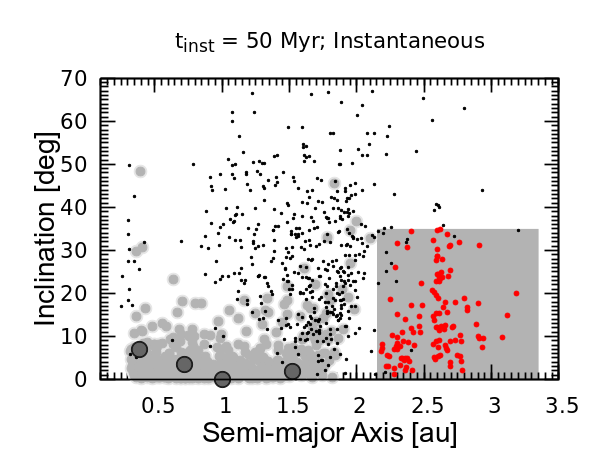}
\includegraphics[scale=0.32]{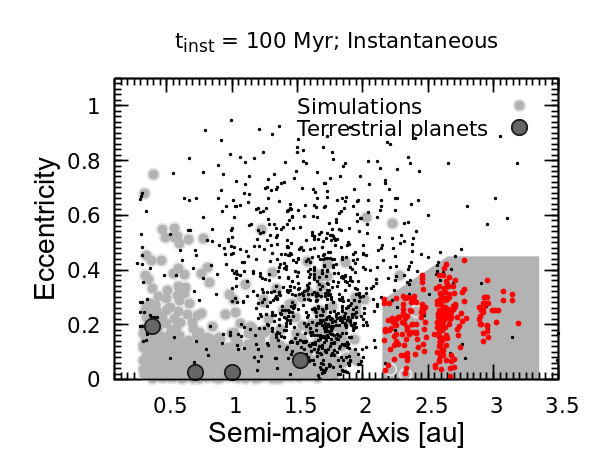}
\includegraphics[scale=0.32]{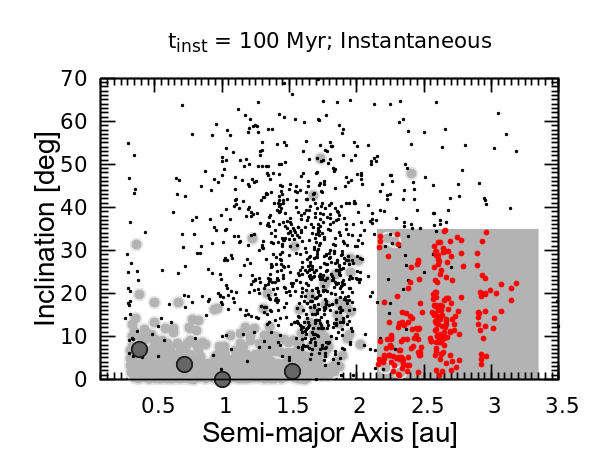}
\caption{Final orbital distributions of all planetary objects at $t=$~200 Myr in four sets of simulations with different instability times. The timing of the instability is indicated at the top of each panel, relative to the time of the gas disk dispersal. Each panel-row show the outcomes of simulations  with different instability times, grouped all together regardless of the ring surface density profile.  Left-panels show  semi-major axis versus eccentricity. Right-panels show semi-major axis versus orbital inclination. Light-grey dots show planets. Small black dots show leftover planetesimals. Red dots show planetesimals (``asteroids'') that have been scattered from the ring and trapped into the asteroid belt. The four dark-grey dots represent the real terrestrial planets. The giant planet dynamical instability is mimicked by assuming that Jupiter and Saturn instantaneously jump from their pre-instability orbits to their current ones at $t=t_{\rm inst}$. The grey-shaded areas show the limits of the asteroid belt.}
    \label{fig:implantation_terrestrial}
\end{figure*}

Figure \ref{fig:implantation_terrestrial} shows the orbital distribution of objects in simulations of Figure \ref{fig:finalplanets}. Unlike the interpolated case -- where we present implanted planetesimals just after the instability --  we now show the distribution of planetesimals in the belt at the end of the process of accretion of the terrestrial planets. As before, planetesimals within the asteroid belt limits are shown as red dots.  From top-to-bottom panels of Figure \ref{fig:implantation_terrestrial}, it is visually clear that the planetesimal implantation efficiency correlates with the timing of the dynamical instability, where the number of implanted objects increases when one goes to later instability times. This is not a surprising result and it is consistent with the results of the previous section where the instability is modeled via numerical interpolation.




Figure \ref{fig:implantation_terrestrial} shows that, in the case where $t_{\rm inst}=0$, about ten planetesimals were implanted in the belt in about 200 simulations. Out of these 10 planetesimals, 3  were implanted in the middle and outer parts of the asteroid belt (top panels).  We have verified that 2 simulations with $t_{\rm inst}=0$ (Figure \ref{fig:implantation_terrestrial}) that implanted at least one planetesimal in the belt also produced reasonable Mars-analogues. However, the the total implantation efficiencies of all our simulations with $t_{\rm inst}=0$, regardless of the ring scenario, are too low to explain the entire population of S-complex type asteroids in the belt today. The required implantation efficiencies to match the S-complex type asteroid population of Table \ref{tab:200Myr} are $\sim4\times10^{-4}$  and  $\sim4\times10^{-3}$. These values are a factor of 5 lower than those of Table \ref{tab:tableafter} because Table \ref{tab:200Myr} simulations  were integrated up to 200 Myr and the asteroid belt depletion during this phase is already accounted for.

Table \ref{tab:200Myr}  shows that even simulations with the shallowest ring slope ($x=0$) -- which tend to produce relatively higher implantation efficiencies --   produce implantation efficiencies about a factor of two lower than that required to account for the total mass carried by S-complex type asteroids combined. We did not perform ``Instantenous''  instability simulations for $t_{\rm inst}=5$~Myr.

For an instability occurring at $t_{\rm inst}=10$~Myr and $x=0$ the implantation efficiency is consistent with the current belt. 
A caveat, however, is that a relatively large number of  ``Mars-analogues'' produced in these specific simulations  tend to be more massive than Mars \citep{izidoroetal15c,izidoroetal22}. In our simulations with $x=0$ about 30\% of the planets between 1.25 and 1.8~au have masses lower than 0.3~$M_{\oplus}$ (see Figure \ref{fig:beforeinstability}). Comparatively, for simulations with  $x=1$, $x=5.5$ and  upside down U-shape disk profiles the relative numbers of Mars-analogues are 40\%, 87\% and 40\%, respectively.  

\begin{figure}[!h]
\centering
\includegraphics[scale=0.39]{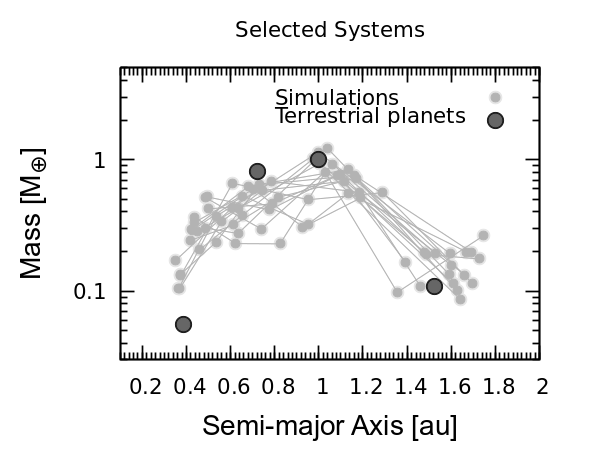}
\includegraphics[scale=0.39]{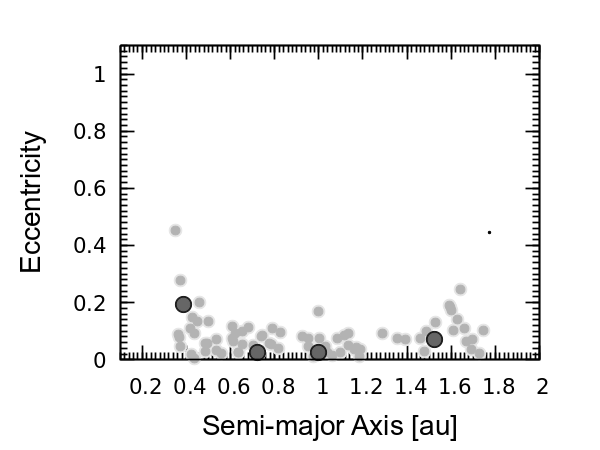}
\includegraphics[scale=0.38]{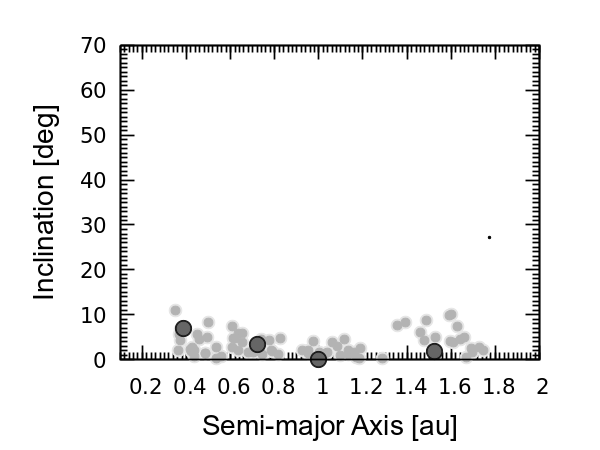}
\caption{Final distributions of solar system analogues produced in selected simulations at $t=$~200 Myr. We only show systems produced in simulations starting with surface density profiles proportional to $r^0$ and $r^{-1}$. The time of the instability is in all shown cases at least 10~Myr.  The x-axes show planets' semi-major axis.  The y-axes show from the top-to-bottom the planet mass, orbital eccentricity and inclination. Light-grey dots show planets produced in our simulations. The four dark-grey dots represent the real terrestrial planets. The giant planet dynamical instability is mimicked by assuming that Jupiter and Saturn instantaneously jump from their pre-instability orbits to their current ones at $t=t_{\rm inst}$.}
    \label{fig:analogues_r0_r1}
\end{figure}

Regardless of the ring radial surface density slope, we  expect that the  instability itself can also help to further deplete the region beyond Mars (as the ring spreads radially out; see Figure \ref{fig:phase1}) if the instability occurs sufficiently early \citep{clementetal2019}. This may help to improve the number of Mars-analogues produced in shallow disks (x=0 and 1). Figure \ref{fig:analogues_r0_r1} shows some selected examples of solar system analogues produced in our shallow disks. Very steep rings profiles (e.g. $x=5.5$) tend to produce a relatively larger number of Mars analogues, but these systems were not as successful in scattering and implanting planetesimals into the belt if the instability happened at $t_{\rm inst}\leq10$~Myr. A ring with a steep  radial profile (e.g. $x=5.5$) would require an instability ocurring at $50$ Myr, at least. Table  \ref{tab:200Myr}  shows that regardless of the planetesimal ring slope,   if the  ``instantaenous'' instability occurs at $\sim$50-100~Myr, planetesimals are implanted in the asteroid belt with total implantation efficiencies consistent with the mass carried by S-complex type asteroids in the belt.

From  results of Tables \ref{tab:tableafter} and \ref{tab:200Myr} altogether, we argue that an instability occurring at $t \gtrsim 5-10$~Myr  relative to the gas disk is necessary to implant the total mass carried by S-complex type asteroids in the belt in the empty asteroid belt formation scenario. Depending on the ring slope this time may have to be pushed to later times.

Our results show that using the implantation efficiency of planetesimals from the terrestrial region into the belt 
to strongly constrain the giant planet instability time remains a challenging exercise, given our limited understanding of planetesimal formation and radial profile of planetesimal rings.




\begin{table}
\vspace{1cm}
\centering
\scriptsize
\caption{Implantation efficiency of planetesimals at the end of simulations of the accretion of terrestrial planets with different ring surface density profiles and instability times. In these simulations the giant planet dynamical instability is modelled as an ``instantaneous jump''  in planetary orbits. The columns are ring slope, instability time (Myr), and implantation efficiency in the inner, middle, and outer belt. The last column gives the total implantation efficiency. Each table-row gives the results of 50 numerical simulations over which efficiencies are computed from\label{tab:200Myr}.}
\begin{tabular}{c c c c c c}
\hline
\multicolumn{1}{c}{}   &   \multicolumn{5}{c}{Implantation efficiency after 200 Myr}           \\
\multicolumn{1}{c}{Ring}  & \multicolumn{1}{c}{Inst.} &    \multicolumn{3}{c}{}         \\
   slope                   &       time (Myr)                       &                     \multicolumn{1}{c}{Inner}  & \multicolumn{1}{c}{Midlle}  & \multicolumn{1}{c}{Outer}& \multicolumn{1}{c}{Total}\\
   
      \hline
0         &   0    &   1.96e-4    &    3.93e-5   &     $<$3.93e-5        &   2.35e-4        \\    \hline
1         &   0    &   4.36e-5  &    $<$4.36e-5   &    4.36e-5        &   8.71e-5        \\    \hline
5.5         &   0    &   7.25e-5  &    $<$7.25e-5   &    $<$7.25e-5        &   7.25e-5        \\    \hline
U        &   0    &   1.68e-4  &   $<$5.59e-5   &    $<$5.59e-5        &   1.68e-4        \\  
\hline\hline
0         &   10    &   7.07e-4    &    1.96e-4   &     $<$3.93e-5        &   {\bf 9.04e-4}        \\    \hline
1         &   10    &   1.74e-04    &    $<$4.36e-5   &     $<$4.36e-5        &   1.74e-4        \\    \hline
5.5       &   10    &     $<$7.25e-5   &     $<$7.25e-5   &     $<$7.25e-5        &   $<$7.25e-5      \\    \hline
U       &   10    &     5.59e-5   &     $<$5.59e-5    &     $<$5.59e-5         &   5.59e-5      \\  
\hline \hline
0         &   50    &  5.89e-4    &    1.19e-3   &     3.93e-5        &   {\bf 1.80e-3}        \\    \hline
1         &   50    &  7.84e-4    &    1.22e-3   &     4.36e-5        &   {\bf 2.05e-3}        \\    \hline
5.5         &   50    &  2.17e-4    &   2.90e-4 &      $<$7.25e-5        &    {\bf 5.07e-4}   \\    \hline
U         &   50    &  5.02e-04    &   6.15e-4 &      5.59e-5        &    {\bf 1.17e-3}     \\  
\hline \hline
0         &   100    &  1.75e-3    &    2.11e-3   &     1.17e-4      &  {\bf 3.41e-3}    \\    \hline
1         &   100    & 1.39e-3    &    1.39e-3   &     8.716e-5      &  {\bf 2.88e-3}   \\    \hline
5.5         &   100    & 1.45e-4    &    3.62e-4   &     $<$7.25e-5      &  {\bf 5.07e-4}   \\    \hline
U         &   100    & 5.59e-04    &    7.82e-4   &     1.12e-4      &  {\bf 1.45e-3}   \\    \hline
\end{tabular}
\end{table}

\section{Conclusion}\label{sec:conclusion}

In this work, we have studied the effects of the solar system dynamical instability on the implantation of planetesimals from the terrestrial region into the asteroid belt. These objects are envisioned to represent S-complex type asteroids.  We have carried out a suite of numerical simulations of the accretion of terrestrial planets. We have performed simulations  for rings of planetesimals extending from 0.7 to 1.5~au, with surface density profiles proportional to $r^{-x}$, were x is assumed to be 0, 1, and 5.5. We also studied a scenario with an upside-down U-shape planetesimal ring profile. Our model setup is motivated by recent planet formation models suggesting that the terrestrial planets accreted from rings of planetesimals \citep{izidoroetal22,morbidellietal22}, and that the asteroid belt may have been born empty \citep{hansen09,izidoroetal14a,raymondizidoro17b}. We have computed the implantation efficiency of planetesimals into the asteroid belt for giant planet dynamical instabilities occurring at t = 0, 5, 10, 50, and 100 Myr relative to time of sun's natal disk dispersal.

We conclude that the ``Empty asteroid belt'' model \citep{izidoroetal15b,raymondizidoro17b,izidoroetal22}  is broadly consistent with a dynamical instability of the solar system giant planets ocurring at $t\gtrapprox5-10$ Myr, relative to the timing of the gas disk dispersal. A dynamical instability occurring at or  after this time,  allows the total mass carried by S-complex type asteroids to be implanted in the  asteroid belt. Our results suggest that an instability occurring at the timing of the gas disk dispersal \citep{liuetal22} may not be able to implant enough material in the belt to explain these observed populations of asteroids. This scenario may be particularly challenged by the existence of S-complex type asteroids in the outer main belt \citep[e.g.][]{demeocarry14}. The overall low implantation efficiency yielded in this scenario represents a major challenge for a giant planet instability coincident with the gas disk dispersal \citep{liuetal22}, unless the gas disk in the terrestrial region ($<5$au) have vanished at least a few Myr earlier than that in the giant planet region  ($>5$~au). If this is the case, it may give enough time for planetesimals from the terrestrial region to be scattered into the asteroid belt before the giant planet instability phase. This scenario may imply that the solar system was once a long-lived transition disk~\citep[e.g.][]{muzerolleetal10}. Alternatively, a dynamical instability occurring at the timing of the gas disk dispersal may also require that the asteroid belt was not born completely empty.


\section*{Acknowledgments}
 The work of R.D. was supported by the NASA Emerging Worlds program, grant 80NSSC21K0387. MSC is supported by NASA Emerging Worlds grant 80NSSC23K0868 and NASA’s CHAMPs team, supported by NASA under Grant No. 80NSSC21K0905 issued through the Interdisciplinary Consortia for Astrobiology Research (ICAR) program. SNR thanks the CNRS's PNP and MITI/80PRIME programs for support.

\section*{Data Availability}
Simulations of this work are available from the corresponding author on reasonable request.

%



\bibliographystyle{cas-model2-names}
\bibliography{mybib}

\begin{thebibliography}{73}
\expandafter\ifx\csname natexlab\endcsname\relax\def\natexlab#1{#1}\fi
\providecommand{\url}[1]{\texttt{#1}}
\providecommand{\href}[2]{#2}
\providecommand{\path}[1]{#1}
\providecommand{\DOIprefix}{doi:}
\providecommand{\ArXivprefix}{arXiv:}
\providecommand{\URLprefix}{URL: }
\providecommand{\Pubmedprefix}{pmid:}
\providecommand{\doi}[1]{\href{http://dx.doi.org/#1}{\path{#1}}}
\providecommand{\Pubmed}[1]{\href{pmid:#1}{\path{#1}}}
\providecommand{\bibinfo}[2]{#2}
\ifx\xfnm\relax \def\xfnm[#1]{\unskip,\space#1}\fi
\bibitem[{Agnor(1999)}]{agnor99}
\bibinfo{author}{Agnor, C.}, \bibinfo{year}{1999}.
\newblock \bibinfo{title}{{On the Character and Consequences of Large Impacts
  in the Late Stage of Terrestrial Planet Formation}}.
\newblock \bibinfo{journal}{Icarus} \bibinfo{volume}{142},
  \bibinfo{pages}{219--237}.
\newblock \URLprefix \url{http://dx.doi.org/10.1006/icar.1999.6201},
  \DOIprefix\doi{10.1006/icar.1999.6201}.
\bibitem[{{Bottke} et~al.(2006){Bottke}, {Nesvorn{\'y}}, {Grimm}, {Morbidelli}
  and {O'Brien}}]{bottkeetal06}
\bibinfo{author}{{Bottke}, W.F.}, \bibinfo{author}{{Nesvorn{\'y}}, D.},
  \bibinfo{author}{{Grimm}, R.E.}, \bibinfo{author}{{Morbidelli}, A.},
  \bibinfo{author}{{O'Brien}, D.P.}, \bibinfo{year}{2006}.
\newblock \bibinfo{title}{{Iron meteorites as remnants of planetesimals formed
  in the terrestrial planet region}}.
\newblock \bibinfo{journal}{\nat} \bibinfo{volume}{439},
  \bibinfo{pages}{821--824}.
\newblock \DOIprefix\doi{10.1038/nature04536}.
\bibitem[{Bottke et~al.(2012)Bottke, Vokrouhlick{\'{y}}, Minton,
  Nesvorn{\'{y}}, Morbidelli, Brasser, Simonson and Levison}]{bottkeetal12}
\bibinfo{author}{Bottke, W.F.}, \bibinfo{author}{Vokrouhlick{\'{y}}, D.},
  \bibinfo{author}{Minton, D.}, \bibinfo{author}{Nesvorn{\'{y}}, D.},
  \bibinfo{author}{Morbidelli, A.}, \bibinfo{author}{Brasser, R.},
  \bibinfo{author}{Simonson, B.}, \bibinfo{author}{Levison, H.F.},
  \bibinfo{year}{2012}.
\newblock \bibinfo{title}{{An Archaean heavy bombardment from a destabilized
  extension of the asteroid belt}}.
\newblock \bibinfo{journal}{Nature} \bibinfo{volume}{485},
  \bibinfo{pages}{78--81}.
\newblock \DOIprefix\doi{10.1038/nature10967}.
\bibitem[{Brasil et~al.(2016)Brasil, Roig, Nesvorn{\'{y}}, Carruba, Aljbaae and
  Huaman}]{brasiletal16}
\bibinfo{author}{Brasil, P.}, \bibinfo{author}{Roig, F.},
  \bibinfo{author}{Nesvorn{\'{y}}, D.}, \bibinfo{author}{Carruba, V.},
  \bibinfo{author}{Aljbaae, S.}, \bibinfo{author}{Huaman, M.},
  \bibinfo{year}{2016}.
\newblock \bibinfo{title}{{Dynamical dispersal of primordial asteroid
  families}}.
\newblock \bibinfo{journal}{Icarus} \bibinfo{volume}{266},
  \bibinfo{pages}{142--151}.
\newblock \URLprefix \url{http://adsabs.harvard.edu/abs/2016Icar..266..142B},
  \DOIprefix\doi{10.1016/j.icarus.2015.11.015}.
\bibitem[{Brasser et~al.(2007)Brasser, Duncan and Levison}]{brasseretal07}
\bibinfo{author}{Brasser, R.}, \bibinfo{author}{Duncan, M.},
  \bibinfo{author}{Levison, H.}, \bibinfo{year}{2007}.
\newblock \bibinfo{title}{{Embedded star clusters and the formation of the Oort
  cloud}}.
\newblock \bibinfo{journal}{Icarus} \bibinfo{volume}{191},
  \bibinfo{pages}{413--433}.
\newblock \URLprefix \url{http://adsabs.harvard.edu/abs/2007Icar..191..413B
  http://linkinghub.elsevier.com/retrieve/pii/S0019103507002205},
  \DOIprefix\doi{10.1016/j.icarus.2007.05.003}.
\bibitem[{Budde et~al.(2016)Budde, Burkhardt, Brennecka, Fischer-Gödde,
  Kruijer and Kleine}]{buddeetal2016}
\bibinfo{author}{Budde, G.}, \bibinfo{author}{Burkhardt, C.},
  \bibinfo{author}{Brennecka, G.A.}, \bibinfo{author}{Fischer-Gödde, M.},
  \bibinfo{author}{Kruijer, T.S.}, \bibinfo{author}{Kleine, T.},
  \bibinfo{year}{2016}.
\newblock \bibinfo{title}{Molybdenum isotopic evidence for the origin of
  chondrules and a distinct genetic heritage of carbonaceous and
  non-carbonaceous meteorites}.
\newblock \bibinfo{journal}{Earth and Planetary Science Letters}
  \bibinfo{volume}{454}, \bibinfo{pages}{293--303}.
\newblock \URLprefix
  \url{https://www.sciencedirect.com/science/article/pii/S0012821X16305003},
  \DOIprefix\doi{https://doi.org/10.1016/j.epsl.2016.09.020}.
\bibitem[{Chambers(1999)}]{chambers99}
\bibinfo{author}{Chambers, J.E.}, \bibinfo{year}{1999}.
\newblock \bibinfo{title}{{A hybrid symplectic integrator that permits close
  encounters between massive bodies}}.
\newblock \bibinfo{journal}{\mnras} \bibinfo{volume}{304},
  \bibinfo{pages}{793--799}.
\newblock \URLprefix
  \url{http://mnras.oxfordjournals.org/cgi/doi/10.1046/j.1365-8711.1999.02379.x$\backslash$nhttp://adsabs.harvard.edu/abs/1999MNRAS.304..793C$\backslash$nhttp://adsabs.harvard.edu/cgi-bin/nph-data{\_}query?bibcode=1999MNRAS.304..793C{\&}link{\_}type=ARTICLE},
  \DOIprefix\doi{10.1046/j.1365-8711.1999.02379.x}.
\bibitem[{Chambers(2001)}]{chambers01}
\bibinfo{author}{Chambers, J.E.}, \bibinfo{year}{2001}.
\newblock \bibinfo{title}{{Making More Terrestrial Planets}}.
\newblock \bibinfo{journal}{Icarus} \bibinfo{volume}{152},
  \bibinfo{pages}{205}.
\newblock \URLprefix
  \url{http://adsabs.harvard.edu/cgi-bin/nph-data{\_}query?bibcode=2001Icar..152..205C{\&}link{\_}type=ABSTRACT$\backslash$npapers://0be24a46-325a-4116-a3c6-fd8a3b614472/Paper/p12938},
  \DOIprefix\doi{10.1006/icar.2001.6639}.
\bibitem[{{Clement} et~al.(2018){Clement}, {Kaib}, {Raymond} and
  {Walsh}}]{clementetal18}
\bibinfo{author}{{Clement}, M.S.}, \bibinfo{author}{{Kaib}, N.A.},
  \bibinfo{author}{{Raymond}, S.N.}, \bibinfo{author}{{Walsh}, K.J.},
  \bibinfo{year}{2018}.
\newblock \bibinfo{title}{{Mars' growth stunted by an early giant planet
  instability}}.
\newblock \bibinfo{journal}{\icarus} \bibinfo{volume}{311},
  \bibinfo{pages}{340--356}.
\newblock \DOIprefix\doi{10.1016/j.icarus.2018.04.008},
  \href{http://arxiv.org/abs/1804.04233}{\tt arXiv:1804.04233}.
\bibitem[{Clement et~al.(2019)Clement, Raymond and Kaib}]{clementetal2019}
\bibinfo{author}{Clement, M.S.}, \bibinfo{author}{Raymond, S.N.},
  \bibinfo{author}{Kaib, N.A.}, \bibinfo{year}{2019}.
\newblock \bibinfo{title}{Excitation and depletion of the asteroid belt in the
  early instability scenario}.
\newblock \bibinfo{journal}{The Astronomical Journal} \bibinfo{volume}{157},
  \bibinfo{pages}{38}.
\newblock \URLprefix \url{https://dx.doi.org/10.3847/1538-3881/aaf21e},
  \DOIprefix\doi{10.3847/1538-3881/aaf21e}.
\bibitem[{{Clement} et~al.(2021){Clement}, {Raymond}, {Kaib}, {Deienno},
  {Chambers} and {Izidoro}}]{clementetal21}
\bibinfo{author}{{Clement}, M.S.}, \bibinfo{author}{{Raymond}, S.N.},
  \bibinfo{author}{{Kaib}, N.A.}, \bibinfo{author}{{Deienno}, R.},
  \bibinfo{author}{{Chambers}, J.E.}, \bibinfo{author}{{Izidoro}, A.},
  \bibinfo{year}{2021}.
\newblock \bibinfo{title}{{Born eccentric: Constraints on Jupiter and Saturn's
  pre-instability orbits}}.
\newblock \bibinfo{journal}{\icarus} \bibinfo{volume}{355},
  \bibinfo{pages}{114122}.
\newblock \DOIprefix\doi{10.1016/j.icarus.2020.114122},
  \href{http://arxiv.org/abs/2009.11323}{\tt arXiv:2009.11323}.
\bibitem[{Deienno et~al.(2016)Deienno, Gomes, Walsh, Morbidelli and
  Nesvorn{\'{y}}}]{deiennoetal16}
\bibinfo{author}{Deienno, R.}, \bibinfo{author}{Gomes, R.S.},
  \bibinfo{author}{Walsh, K.J.}, \bibinfo{author}{Morbidelli, A.},
  \bibinfo{author}{Nesvorn{\'{y}}, D.}, \bibinfo{year}{2016}.
\newblock \bibinfo{title}{{Is the Grand Tack model compatible with the orbital
  distribution of main belt asteroids?}}
\newblock \bibinfo{journal}{Icarus} \bibinfo{volume}{272},
  \bibinfo{pages}{114--124}.
\newblock \URLprefix
  \url{http://linkinghub.elsevier.com/retrieve/pii/S0019103516001214},
  \DOIprefix\doi{10.1016/j.icarus.2016.02.043}.
\bibitem[{{Deienno} et~al.(2018){Deienno}, {Izidoro}, {Morbidelli}, {Gomes},
  {Nesvorn{\'y}} and {Raymond}}]{deiennoetal18}
\bibinfo{author}{{Deienno}, R.}, \bibinfo{author}{{Izidoro}, A.},
  \bibinfo{author}{{Morbidelli}, A.}, \bibinfo{author}{{Gomes}, R.S.},
  \bibinfo{author}{{Nesvorn{\'y}}, D.}, \bibinfo{author}{{Raymond}, S.N.},
  \bibinfo{year}{2018}.
\newblock \bibinfo{title}{{Excitation of a Primordial Cold Asteroid Belt as an
  Outcome of Planetary Instability}}.
\newblock \bibinfo{journal}{\apj} \bibinfo{volume}{864}, \bibinfo{pages}{50}.
\newblock \DOIprefix\doi{10.3847/1538-4357/aad55d},
  \href{http://arxiv.org/abs/1808.00609}{\tt arXiv:1808.00609}.
\bibitem[{{Deienno} et~al.(2017){Deienno}, {Morbidelli}, {Gomes} and
  {Nesvorn{\'y}}}]{deiennoetal17}
\bibinfo{author}{{Deienno}, R.}, \bibinfo{author}{{Morbidelli}, A.},
  \bibinfo{author}{{Gomes}, R.S.}, \bibinfo{author}{{Nesvorn{\'y}}, D.},
  \bibinfo{year}{2017}.
\newblock \bibinfo{title}{{Constraining the Giant Planets{\textquoteright}
  Initial Configuration from Their Evolution: Implications for the Timing of
  the Planetary Instability}}.
\newblock \bibinfo{journal}{\aj} \bibinfo{volume}{153}, \bibinfo{pages}{153}.
\newblock \DOIprefix\doi{10.3847/1538-3881/aa5eaa},
  \href{http://arxiv.org/abs/1702.02094}{\tt arXiv:1702.02094}.
\bibitem[{{Deienno} et~al.(2019){Deienno}, {Walsh}, {Kretke} and
  {Levison}}]{deiennoetal19}
\bibinfo{author}{{Deienno}, R.}, \bibinfo{author}{{Walsh}, K.J.},
  \bibinfo{author}{{Kretke}, K.A.}, \bibinfo{author}{{Levison}, H.F.},
  \bibinfo{year}{2019}.
\newblock \bibinfo{title}{{Energy Dissipation in Large
  Collisions{\textemdash}No Change in Planet Formation Outcomes}}.
\newblock \bibinfo{journal}{\apj} \bibinfo{volume}{876}, \bibinfo{pages}{103}.
\newblock \DOIprefix\doi{10.3847/1538-4357/ab16e1}.
\bibitem[{DeMeo and Carry(2014)}]{demeocarry14}
\bibinfo{author}{DeMeo, F.E.}, \bibinfo{author}{Carry, B.},
  \bibinfo{year}{2014}.
\newblock \bibinfo{title}{{Solar System evolution from compositional mapping of
  the asteroid belt.}}
\newblock \bibinfo{journal}{Nature} \bibinfo{volume}{505},
  \bibinfo{pages}{629--34}.
\newblock \URLprefix \url{http://dx.doi.org/10.1038/nature12908},
  \DOIprefix\doi{10.1038/nature12908},
  \href{http://arxiv.org/abs/1408.2787}{\tt arXiv:1408.2787}.
\bibitem[{{Dr{\c a}{\.z}kowska} and {Alibert}(2017)}]{drazkowskaalibert17}
\bibinfo{author}{{Dr{\c a}{\.z}kowska}, J.}, \bibinfo{author}{{Alibert}, Y.},
  \bibinfo{year}{2017}.
\newblock \bibinfo{title}{{Planetesimal formation starts at the snow line}}.
\newblock \bibinfo{journal}{\aap} \bibinfo{volume}{608}, \bibinfo{pages}{A92}.
\newblock \DOIprefix\doi{10.1051/0004-6361/201731491},
  \href{http://arxiv.org/abs/1710.00009}{\tt arXiv:1710.00009}.
\bibitem[{{Gerbig} et~al.(2019){Gerbig}, {Lenz} and {Klahr}}]{gerbigetal19}
\bibinfo{author}{{Gerbig}, K.}, \bibinfo{author}{{Lenz}, C.T.},
  \bibinfo{author}{{Klahr}, H.}, \bibinfo{year}{2019}.
\newblock \bibinfo{title}{{Linking planetesimal and dust content in
  protoplanetary disks via a local toy model}}.
\newblock \bibinfo{journal}{\aap} \bibinfo{volume}{629}, \bibinfo{pages}{A116}.
\newblock \DOIprefix\doi{10.1051/0004-6361/201935278},
  \href{http://arxiv.org/abs/1908.02608}{\tt arXiv:1908.02608}.
\bibitem[{Gomes et~al.(2005)Gomes, Levison, Tsiganis and
  Morbidelli}]{gomesetal05}
\bibinfo{author}{Gomes, R.}, \bibinfo{author}{Levison, H.F.},
  \bibinfo{author}{Tsiganis, K.}, \bibinfo{author}{Morbidelli, A.},
  \bibinfo{year}{2005}.
\newblock \bibinfo{title}{{Origin of the cataclysmic Late Heavy Bombardment
  period of the terrestrial planets.}}
\newblock \bibinfo{journal}{Nature} \bibinfo{volume}{435},
  \bibinfo{pages}{466--9}.
\newblock \URLprefix \url{http://dx.doi.org/10.1038/nature03676},
  \DOIprefix\doi{10.1038/nature03676}.
\bibitem[{{Gomes} et~al.(2018){Gomes}, {Nesvorn{\'y}}, {Morbidelli}, {Deienno}
  and {Nogueira}}]{gomesetal18}
\bibinfo{author}{{Gomes}, R.}, \bibinfo{author}{{Nesvorn{\'y}}, D.},
  \bibinfo{author}{{Morbidelli}, A.}, \bibinfo{author}{{Deienno}, R.},
  \bibinfo{author}{{Nogueira}, E.}, \bibinfo{year}{2018}.
\newblock \bibinfo{title}{{Checking the compatibility of the cold Kuiper belt
  with a planetary instability migration model}}.
\newblock \bibinfo{journal}{\icarus} \bibinfo{volume}{306},
  \bibinfo{pages}{319--327}.
\newblock \DOIprefix\doi{10.1016/j.icarus.2017.10.018},
  \href{http://arxiv.org/abs/1710.05178}{\tt arXiv:1710.05178}.
\bibitem[{Gradie and Tedesco(1982)}]{gradietedesco82}
\bibinfo{author}{Gradie, J.}, \bibinfo{author}{Tedesco, E.},
  \bibinfo{year}{1982}.
\newblock \bibinfo{title}{{Compositional structure of the asteroid belt.}}
\newblock \bibinfo{journal}{Science} \bibinfo{volume}{216},
  \bibinfo{pages}{1405--1407}.
\newblock \URLprefix
  \url{http://science.sciencemag.org/content/216/4553/1405.abstract},
  \DOIprefix\doi{10.1126/science.216.4553.1405}.
\bibitem[{{Grewal} et~al.(2021){Grewal}, {Dasgupta} and {Marty}}]{grewaletal21}
\bibinfo{author}{{Grewal}, D.S.}, \bibinfo{author}{{Dasgupta}, R.},
  \bibinfo{author}{{Marty}, B.}, \bibinfo{year}{2021}.
\newblock \bibinfo{title}{{A very early origin of isotopically distinct
  nitrogen in inner Solar System protoplanets}}.
\newblock \bibinfo{journal}{Nature Astronomy}
  \DOIprefix\doi{10.1038/s41550-020-01283-y}.
\bibitem[{Hansen(2009)}]{hansen09}
\bibinfo{author}{Hansen, B.M.S.}, \bibinfo{year}{2009}.
\newblock \bibinfo{title}{{Formation of the Terrestrial Planets From a Narrow
  Annulus}}.
\newblock \bibinfo{journal}{\apj} \bibinfo{volume}{703},
  \bibinfo{pages}{1131--1140}.
\newblock \URLprefix \url{http://adsabs.harvard.edu/abs/2009ApJ...703.1131H},
  \DOIprefix\doi{10.1088/0004-637X/703/1/1131},
  \href{http://arxiv.org/abs/0908.0743}{\tt arXiv:0908.0743}.
\bibitem[{{Izidoro} et~al.(2021){Izidoro}, {Bitsch} and
  {Dasgupta}}]{izidoroetal21}
\bibinfo{author}{{Izidoro}, A.}, \bibinfo{author}{{Bitsch}, B.},
  \bibinfo{author}{{Dasgupta}, R.}, \bibinfo{year}{2021}.
\newblock \bibinfo{title}{{The effect of a strong pressure bump in the Sun's
  natal disk: Terrestrial planet formation via planetesimal accretion rather
  than pebble accretion}}.
\newblock \bibinfo{journal}{arXiv e-prints} ,
  \bibinfo{pages}{arXiv:2105.01101}\href{http://arxiv.org/abs/2105.01101}{\tt
  arXiv:2105.01101}.
\bibitem[{{Izidoro} et~al.(2022){Izidoro}, {Dasgupta}, {Raymond}, {Deienno},
  {Bitsch} and {Isella}}]{izidoroetal22}
\bibinfo{author}{{Izidoro}, A.}, \bibinfo{author}{{Dasgupta}, R.},
  \bibinfo{author}{{Raymond}, S.N.}, \bibinfo{author}{{Deienno}, R.},
  \bibinfo{author}{{Bitsch}, B.}, \bibinfo{author}{{Isella}, A.},
  \bibinfo{year}{2022}.
\newblock \bibinfo{title}{{Planetesimal rings as the cause of the Solar
  System's planetary architecture}}.
\newblock \bibinfo{journal}{Nature Astronomy} \bibinfo{volume}{6},
  \bibinfo{pages}{357--366}.
\newblock \DOIprefix\doi{10.1038/s41550-021-01557-z},
  \href{http://arxiv.org/abs/2112.15558}{\tt arXiv:2112.15558}.
\bibitem[{Izidoro et~al.(2014)Izidoro, Haghighipour, Winter and
  Tsuchida}]{izidoroetal14a}
\bibinfo{author}{Izidoro, A.}, \bibinfo{author}{Haghighipour, N.},
  \bibinfo{author}{Winter, O.C.}, \bibinfo{author}{Tsuchida, M.},
  \bibinfo{year}{2014}.
\newblock \bibinfo{title}{{Terrestrial Planet Formation in a Protoplanetary
  Disk With a Local Mass Depletion: a Successful Scenario for the Formation of
  Mars}}.
\newblock \bibinfo{journal}{\apj} \bibinfo{volume}{782}, \bibinfo{pages}{31}.
\newblock \URLprefix
  \url{http://stacks.iop.org/0004-637X/782/i=1/a=31?key=crossref.b3b645d2c2269cc448f613acbe4cc7dc},
  \DOIprefix\doi{10.1088/0004-637X/782/1/31}.
\bibitem[{{Izidoro} et~al.(2015){Izidoro}, {Morbidelli}, {Raymond}, {Hersant}
  and {Pierens}}]{izidoroetal15c}
\bibinfo{author}{{Izidoro}, A.}, \bibinfo{author}{{Morbidelli}, A.},
  \bibinfo{author}{{Raymond}, S.N.}, \bibinfo{author}{{Hersant}, F.},
  \bibinfo{author}{{Pierens}, A.}, \bibinfo{year}{2015}.
\newblock \bibinfo{title}{{Accretion of Uranus and Neptune from
  inward-migrating planetary embryos blocked by Jupiter and Saturn}}.
\newblock \bibinfo{journal}{Astronomy \& Astrophysics} \bibinfo{volume}{582},
  \bibinfo{pages}{A99}.
\newblock \DOIprefix\doi{10.1051/0004-6361/201425525},
  \href{http://arxiv.org/abs/1506.03029}{\tt arXiv:1506.03029}.
\bibitem[{Izidoro and Raymond(2018)}]{izidororaymond18}
\bibinfo{author}{Izidoro, A.}, \bibinfo{author}{Raymond, S.N.},
  \bibinfo{year}{2018}.
\newblock \bibinfo{title}{Formation of Terrestrial Planets}.
  \bibinfo{publisher}{Springer International Publishing},
  \bibinfo{address}{Cham}.
\newblock pp. \bibinfo{pages}{1--59}.
\newblock \URLprefix \url{https://doi.org/10.1007/978-3-319-30648-3_142-1},
  \DOIprefix\doi{10.1007/978-3-319-30648-3_142-1}.
\bibitem[{Izidoro et~al.(2015)Izidoro, Raymond, Morbidelli and
  Winter}]{izidoroetal15b}
\bibinfo{author}{Izidoro, A.}, \bibinfo{author}{Raymond, S.N.},
  \bibinfo{author}{Morbidelli, A.}, \bibinfo{author}{Winter, O.C.},
  \bibinfo{year}{2015}.
\newblock \bibinfo{title}{{Terrestrial planet formation constrained by Mars and
  the structure of the asteroid belt}}.
\newblock \bibinfo{journal}{\mnras} \bibinfo{volume}{453},
  \bibinfo{pages}{3619--3634}.
\newblock \URLprefix
  \url{http://mnras.oxfordjournals.org/lookup/doi/10.1093/mnras/stv1835},
  \DOIprefix\doi{10.1093/mnras/stv1835},
  \href{http://arxiv.org/abs/1508.01365}{\tt arXiv:1508.01365}.
\bibitem[{{Izidoro} et~al.(2016){Izidoro}, {Raymond}, {Pierens}, {Morbidelli},
  {Winter} and {Nesvorny`}}]{izidoroetal16}
\bibinfo{author}{{Izidoro}, A.}, \bibinfo{author}{{Raymond}, S.N.},
  \bibinfo{author}{{Pierens}, A.}, \bibinfo{author}{{Morbidelli}, A.},
  \bibinfo{author}{{Winter}, O.C.}, \bibinfo{author}{{Nesvorny`}, D.},
  \bibinfo{year}{2016}.
\newblock \bibinfo{title}{{The Asteroid Belt as a Relic from a Chaotic Early
  Solar System}}.
\newblock \bibinfo{journal}{\apj} \bibinfo{volume}{833}, \bibinfo{pages}{40}.
\newblock \DOIprefix\doi{10.3847/1538-4357/833/1/40},
  \href{http://arxiv.org/abs/1609.04970}{\tt arXiv:1609.04970}.
\bibitem[{{Kaib} and {Chambers}(2016)}]{kaibchambers16}
\bibinfo{author}{{Kaib}, N.A.}, \bibinfo{author}{{Chambers}, J.E.},
  \bibinfo{year}{2016}.
\newblock \bibinfo{title}{{The fragility of the terrestrial planets during a
  giant-planet instability}}.
\newblock \bibinfo{journal}{\mnras} \bibinfo{volume}{455},
  \bibinfo{pages}{3561--3569}.
\newblock \DOIprefix\doi{10.1093/mnras/stv2554},
  \href{http://arxiv.org/abs/1510.08448}{\tt arXiv:1510.08448}.
\bibitem[{Kleine et~al.(2009)Kleine, Touboul, Bourdon, Nimmo, Mezger, Palme,
  Jacobsen, Yin and Halliday}]{kleinetal09}
\bibinfo{author}{Kleine, T.}, \bibinfo{author}{Touboul, M.},
  \bibinfo{author}{Bourdon, B.}, \bibinfo{author}{Nimmo, F.},
  \bibinfo{author}{Mezger, K.}, \bibinfo{author}{Palme, H.},
  \bibinfo{author}{Jacobsen, S.B.}, \bibinfo{author}{Yin, Q.Z.},
  \bibinfo{author}{Halliday, A.N.}, \bibinfo{year}{2009}.
\newblock \bibinfo{title}{{Hf-W chronology of the accretion and early evolution
  of asteroids and terrestrial planets}}.
\newblock \bibinfo{journal}{Geochim. Cosmochim. Acta} \bibinfo{volume}{73},
  \bibinfo{pages}{5150--5188}.
\newblock \URLprefix
  \url{http://linkinghub.elsevier.com/retrieve/pii/S0016703709003287},
  \DOIprefix\doi{10.1016/j.gca.2008.11.047}.
\bibitem[{Kokubo and Ida(2000)}]{kokuboida00}
\bibinfo{author}{Kokubo, E.}, \bibinfo{author}{Ida, S.}, \bibinfo{year}{2000}.
\newblock \bibinfo{title}{{Formation of Protoplanets from Planetesimals in the
  Solar Nebula}}.
\newblock \bibinfo{journal}{Icarus} \bibinfo{volume}{143}, \bibinfo{pages}{15}.
\newblock \URLprefix
  \url{http://adsabs.harvard.edu/cgi-bin/nph-data{\_}query?bibcode=2000Icar..143...15K{\&}link{\_}type=ABSTRACT$\backslash$npapers://0be24a46-325a-4116-a3c6-fd8a3b614472/Paper/p96},
  \DOIprefix\doi{10.1006/icar.1999.6237}.
\bibitem[{Kruijer et~al.(2017)Kruijer, Burkhardt, Budde and
  Kleine}]{Kruijeretal17}
\bibinfo{author}{Kruijer, T.S.}, \bibinfo{author}{Burkhardt, C.},
  \bibinfo{author}{Budde, G.}, \bibinfo{author}{Kleine, T.},
  \bibinfo{year}{2017}.
\newblock \bibinfo{title}{Age of jupiter inferred from the distinct genetics
  and formation times of meteorites}.
\newblock \bibinfo{journal}{Proceedings of the National Academy of Sciences}
  \bibinfo{volume}{114}, \bibinfo{pages}{6712--6716}.
\newblock \URLprefix \url{https://www.pnas.org/content/114/26/6712},
  \DOIprefix\doi{10.1073/pnas.1704461114},
  \href{http://arxiv.org/abs/https://www.pnas.org/content/114/26/6712.full.pdf}{\tt
  arXiv:https://www.pnas.org/content/114/26/6712.full.pdf}.
\bibitem[{{Kruijer} et~al.(2020){Kruijer}, {Kleine} and {Borg}}]{kruijeretal20}
\bibinfo{author}{{Kruijer}, T.S.}, \bibinfo{author}{{Kleine}, T.},
  \bibinfo{author}{{Borg}, L.E.}, \bibinfo{year}{2020}.
\newblock \bibinfo{title}{{The great isotopic dichotomy of the early Solar
  System}}.
\newblock \bibinfo{journal}{Nature Astronomy} \bibinfo{volume}{4},
  \bibinfo{pages}{32--40}.
\newblock \DOIprefix\doi{10.1038/s41550-019-0959-9}.
\bibitem[{Levison et~al.(2009)Levison, Bottke, Gounelle, Morbidelli,
  Nesvorn{\'{y}} and Tsiganis}]{levisonetal09}
\bibinfo{author}{Levison, H.F.}, \bibinfo{author}{Bottke, W.F.},
  \bibinfo{author}{Gounelle, M.}, \bibinfo{author}{Morbidelli, A.},
  \bibinfo{author}{Nesvorn{\'{y}}, D.}, \bibinfo{author}{Tsiganis, K.},
  \bibinfo{year}{2009}.
\newblock \bibinfo{title}{{Contamination of the asteroid belt by primordial
  trans-Neptunian objects.}}
\newblock \bibinfo{journal}{Nature} \bibinfo{volume}{460},
  \bibinfo{pages}{364--6}.
\newblock \URLprefix \url{http://dx.doi.org/10.1038/nature08094},
  \DOIprefix\doi{10.1038/nature08094}.
\bibitem[{{Levison} et~al.(2012){Levison}, {Duncan} and
  {Thommes}}]{levisonetal12}
\bibinfo{author}{{Levison}, H.F.}, \bibinfo{author}{{Duncan}, M.J.},
  \bibinfo{author}{{Thommes}, E.}, \bibinfo{year}{2012}.
\newblock \bibinfo{title}{{A Lagrangian Integrator for Planetary Accretion and
  Dynamics (LIPAD)}}.
\newblock \bibinfo{journal}{\aj} \bibinfo{volume}{144}, \bibinfo{pages}{119}.
\newblock \DOIprefix\doi{10.1088/0004-6256/144/4/119},
  \href{http://arxiv.org/abs/1207.0754}{\tt arXiv:1207.0754}.
\bibitem[{Levison et~al.(2011)Levison, Morbidelli, Tsiganis, Nesvorn{\'{y}} and
  Gomes}]{levisonetal11}
\bibinfo{author}{Levison, H.F.}, \bibinfo{author}{Morbidelli, A.},
  \bibinfo{author}{Tsiganis, K.}, \bibinfo{author}{Nesvorn{\'{y}}, D.},
  \bibinfo{author}{Gomes, R.}, \bibinfo{year}{2011}.
\newblock \bibinfo{title}{{Late Orbital Instabilities in the Outer Planets
  Induced By Interaction With a Self-Gravitating Planetesimal Disk}}.
\newblock \bibinfo{journal}{Astron. J.} \bibinfo{volume}{142},
  \bibinfo{pages}{152}.
\newblock \URLprefix
  \url{http://stacks.iop.org/1538-3881/142/i=5/a=152?key=crossref.e96913e8dbf638ca3c28d7e6ffda327a},
  \DOIprefix\doi{10.1088/0004-6256/142/5/152}.
\bibitem[{{Liu} et~al.(2022){Liu}, {Raymond} and {Jacobson}}]{liuetal22}
\bibinfo{author}{{Liu}, B.}, \bibinfo{author}{{Raymond}, S.N.},
  \bibinfo{author}{{Jacobson}, S.A.}, \bibinfo{year}{2022}.
\newblock \bibinfo{title}{{Early Solar System instability triggered by
  dispersal of the gaseous disk}}.
\newblock \bibinfo{journal}{\nat} \bibinfo{volume}{604},
  \bibinfo{pages}{643--646}.
\newblock \DOIprefix\doi{10.1038/s41586-022-04535-1},
  \href{http://arxiv.org/abs/2205.02026}{\tt arXiv:2205.02026}.
\bibitem[{Lykawka and Ito(2013)}]{lykawkaito13}
\bibinfo{author}{Lykawka, P.S.}, \bibinfo{author}{Ito, T.},
  \bibinfo{year}{2013}.
\newblock \bibinfo{title}{{Terrestrial Planet Formation During the Migration}}.
\newblock \bibinfo{journal}{\apj} \bibinfo{volume}{65}, \bibinfo{pages}{65}.
\newblock \URLprefix
  \url{http://stacks.iop.org/0004-637X/773/i=1/a=65?key=crossref.be80a2d8c1f8340fb18aa6f7a19581ea},
  \DOIprefix\doi{10.1088/0004-637X/773/1/65},
  \href{http://arxiv.org/abs/1306.3287}{\tt arXiv:1306.3287}.
\bibitem[{Masset and Snellgrove(2001)}]{massetsnellgrove01}
\bibinfo{author}{Masset, F.}, \bibinfo{author}{Snellgrove, M.},
  \bibinfo{year}{2001}.
\newblock \bibinfo{title}{{Reversing type II migration: Resonance trapping of a
  lighter giant protoplanet}}.
\newblock \bibinfo{journal}{\mnras} \bibinfo{volume}{320},
  \bibinfo{pages}{L55--L59}.
\newblock \DOIprefix\doi{10.1046/j.1365-8711.2001.04159.x},
  \href{http://arxiv.org/abs/0101332}{\tt arXiv:0101332}.
\bibitem[{Minton and Malhotra(2009)}]{mintonmalhotra09}
\bibinfo{author}{Minton, D.A.}, \bibinfo{author}{Malhotra, R.},
  \bibinfo{year}{2009}.
\newblock \bibinfo{title}{{A record of planet migration in the main asteroid
  belt.}}
\newblock \bibinfo{journal}{Nature} \bibinfo{volume}{457},
  \bibinfo{pages}{1109--1111}.
\newblock \URLprefix \url{http://dx.doi.org/10.1038/nature07778},
  \DOIprefix\doi{10.1038/nature07778},
  \href{http://arxiv.org/abs/1102.3131}{\tt arXiv:1102.3131}.
\bibitem[{{Morbidelli} et~al.(2022){Morbidelli}, {Bailli{\'e}}, {Batygin},
  {Charnoz}, {Guillot}, {Rubie} and {Kleine}}]{morbidellietal22}
\bibinfo{author}{{Morbidelli}, A.}, \bibinfo{author}{{Bailli{\'e}}, K.},
  \bibinfo{author}{{Batygin}, K.}, \bibinfo{author}{{Charnoz}, S.},
  \bibinfo{author}{{Guillot}, T.}, \bibinfo{author}{{Rubie}, D.C.},
  \bibinfo{author}{{Kleine}, T.}, \bibinfo{year}{2022}.
\newblock \bibinfo{title}{{Contemporary formation of early Solar System
  planetesimals at two distinct radial locations}}.
\newblock \bibinfo{journal}{Nature Astronomy} \bibinfo{volume}{6},
  \bibinfo{pages}{72--79}.
\newblock \DOIprefix\doi{10.1038/s41550-021-01517-7},
  \href{http://arxiv.org/abs/2112.15413}{\tt arXiv:2112.15413}.
\bibitem[{Morbidelli et~al.(2009)Morbidelli, Bottke, Nesvorn{\'{y}} and
  Levison}]{morbidellietal09b}
\bibinfo{author}{Morbidelli, A.}, \bibinfo{author}{Bottke, W.F.},
  \bibinfo{author}{Nesvorn{\'{y}}, D.}, \bibinfo{author}{Levison, H.F.},
  \bibinfo{year}{2009}.
\newblock \bibinfo{title}{{Asteroids were born big}}.
\newblock \bibinfo{journal}{Icarus} \bibinfo{volume}{204},
  \bibinfo{pages}{558--573}.
\newblock \DOIprefix\doi{10.1016/j.icarus.2009.07.011},
  \href{http://arxiv.org/abs/0907.2512}{\tt arXiv:0907.2512}.
\bibitem[{Morbidelli et~al.(2010)Morbidelli, Brasser, Gomes, Levison and
  Tsiganis}]{morbidellietal10}
\bibinfo{author}{Morbidelli, A.}, \bibinfo{author}{Brasser, R.},
  \bibinfo{author}{Gomes, R.}, \bibinfo{author}{Levison, H.F.},
  \bibinfo{author}{Tsiganis, K.}, \bibinfo{year}{2010}.
\newblock \bibinfo{title}{{Evidence From the Asteroid Belt for a Violent Past
  Evolution of Jupiter'S Orbit}}.
\newblock \bibinfo{journal}{Astron. J.} \bibinfo{volume}{140},
  \bibinfo{pages}{1391--1401}.
\newblock \URLprefix
  \url{http://stacks.iop.org/1538-3881/140/i=5/a=1391?key=crossref.e262f8e313f49990efff9f23d5938812},
  \DOIprefix\doi{10.1088/0004-6256/140/5/1391},
  \href{http://arxiv.org/abs/1009.1521}{\tt arXiv:1009.1521}.
\bibitem[{Morbidelli and Crida(2007)}]{morbidellicrida07}
\bibinfo{author}{Morbidelli, A.}, \bibinfo{author}{Crida, A.},
  \bibinfo{year}{2007}.
\newblock \bibinfo{title}{{The dynamics of Jupiter and Saturn in the gaseous
  protoplanetary disk}}.
\newblock \bibinfo{journal}{Icarus} \bibinfo{volume}{191},
  \bibinfo{pages}{158--171}.
\newblock \URLprefix \url{http://adsabs.harvard.edu/abs/2007Icar..191..158M},
  \DOIprefix\doi{10.1016/j.icarus.2007.04.001},
  \href{http://arxiv.org/abs/0704.1210}{\tt arXiv:0704.1210}.
\bibitem[{Morbidelli and Henrard(1991)}]{morbidellihenrard91}
\bibinfo{author}{Morbidelli, A.}, \bibinfo{author}{Henrard, J.},
  \bibinfo{year}{1991}.
\newblock \bibinfo{title}{{The main secular resonances v6, v5 and v16 in the
  asteroid belt}}.
\newblock \bibinfo{journal}{Celest. Mech. Dyn. Astron.} \bibinfo{volume}{51},
  \bibinfo{pages}{169--197}.
\newblock \URLprefix \url{http://link.springer.com/10.1007/BF00048607},
  \DOIprefix\doi{10.1007/BF00048607}.
\bibitem[{{Morbidelli} et~al.(2018){Morbidelli}, {Nesvorny}, {Laurenz},
  {Marchi}, {Rubie}, {Elkins-Tanton}, {Wieczorek} and
  {Jacobson}}]{morbidellietal18}
\bibinfo{author}{{Morbidelli}, A.}, \bibinfo{author}{{Nesvorny}, D.},
  \bibinfo{author}{{Laurenz}, V.}, \bibinfo{author}{{Marchi}, S.},
  \bibinfo{author}{{Rubie}, D.C.}, \bibinfo{author}{{Elkins-Tanton}, L.},
  \bibinfo{author}{{Wieczorek}, M.}, \bibinfo{author}{{Jacobson}, S.},
  \bibinfo{year}{2018}.
\newblock \bibinfo{title}{{The timeline of the lunar bombardment: Revisited}}.
\newblock \bibinfo{journal}{\icarus} \bibinfo{volume}{305},
  \bibinfo{pages}{262--276}.
\newblock \DOIprefix\doi{10.1016/j.icarus.2017.12.046},
  \href{http://arxiv.org/abs/1801.03756}{\tt arXiv:1801.03756}.
\bibitem[{{Moth{\'e}-Diniz} et~al.(2003){Moth{\'e}-Diniz}, {Carvano} and
  {Lazzaro}}]{mothe-dinizetal03}
\bibinfo{author}{{Moth{\'e}-Diniz}, T.}, \bibinfo{author}{{Carvano},
  J.M.{\'a}.}, \bibinfo{author}{{Lazzaro}, D.}, \bibinfo{year}{2003}.
\newblock \bibinfo{title}{{Distribution of taxonomic classes in the main belt
  of asteroids}}.
\newblock \bibinfo{journal}{\icarus} \bibinfo{volume}{162},
  \bibinfo{pages}{10--21}.
\newblock \DOIprefix\doi{10.1016/S0019-1035(02)00066-0}.
\bibitem[{{Muzerolle} et~al.(2010){Muzerolle}, {Allen}, {Megeath},
  {Hern{\'a}ndez} and {Gutermuth}}]{muzerolleetal10}
\bibinfo{author}{{Muzerolle}, J.}, \bibinfo{author}{{Allen}, L.E.},
  \bibinfo{author}{{Megeath}, S.T.}, \bibinfo{author}{{Hern{\'a}ndez}, J.},
  \bibinfo{author}{{Gutermuth}, R.A.}, \bibinfo{year}{2010}.
\newblock \bibinfo{title}{{A Spitzer Census of Transitional Protoplanetary
  Disks with AU-scale Inner Holes}}.
\newblock \bibinfo{journal}{\apj} \bibinfo{volume}{708},
  \bibinfo{pages}{1107--1118}.
\newblock \DOIprefix\doi{10.1088/0004-637X/708/2/1107},
  \href{http://arxiv.org/abs/0911.2704}{\tt arXiv:0911.2704}.
\bibitem[{{Nesvorn{\'y}} et~al.(2021){Nesvorn{\'y}}, {Roig} and
  {Deienno}}]{nesvornyetal2021}
\bibinfo{author}{{Nesvorn{\'y}}, D.}, \bibinfo{author}{{Roig}, F.V.},
  \bibinfo{author}{{Deienno}, R.}, \bibinfo{year}{2021}.
\newblock \bibinfo{title}{{The Role of Early Giant-planet Instability in
  Terrestrial Planet Formation}}.
\newblock \bibinfo{journal}{\aj} \bibinfo{volume}{161}, \bibinfo{pages}{50}.
\newblock \DOIprefix\doi{10.3847/1538-3881/abc8ef},
  \href{http://arxiv.org/abs/2012.02323}{\tt arXiv:2012.02323}.
\bibitem[{{Nesvorn{\'y}} et~al.(2023){Nesvorn{\'y}}, {Roig},
  {Vokrouhlick{\'y}}, {Bottke}, {Marchi}, {Morbidelli} and
  {Deienno}}]{nesvornyetal23}
\bibinfo{author}{{Nesvorn{\'y}}, D.}, \bibinfo{author}{{Roig}, F.V.},
  \bibinfo{author}{{Vokrouhlick{\'y}}, D.}, \bibinfo{author}{{Bottke}, W.F.},
  \bibinfo{author}{{Marchi}, S.}, \bibinfo{author}{{Morbidelli}, A.},
  \bibinfo{author}{{Deienno}, R.}, \bibinfo{year}{2023}.
\newblock \bibinfo{title}{{Early bombardment of the moon: Connecting the lunar
  crater record to the terrestrial planet formation}}.
\newblock \bibinfo{journal}{\icarus} \bibinfo{volume}{399},
  \bibinfo{pages}{115545}.
\newblock \DOIprefix\doi{10.1016/j.icarus.2023.115545},
  \href{http://arxiv.org/abs/2303.17736}{\tt arXiv:2303.17736}.
\bibitem[{{Nesvorn{\'y}} et~al.(2018){Nesvorn{\'y}}, {Vokrouhlick{\'y}},
  {Bottke} and {Levison}}]{nesvornyetal18}
\bibinfo{author}{{Nesvorn{\'y}}, D.}, \bibinfo{author}{{Vokrouhlick{\'y}}, D.},
  \bibinfo{author}{{Bottke}, W.F.}, \bibinfo{author}{{Levison}, H.F.},
  \bibinfo{year}{2018}.
\newblock \bibinfo{title}{{Evidence for very early migration of the Solar
  System planets from the Patroclus-Menoetius binary Jupiter Trojan}}.
\newblock \bibinfo{journal}{Nature Astronomy} \bibinfo{volume}{2},
  \bibinfo{pages}{878--882}.
\newblock \DOIprefix\doi{10.1038/s41550-018-0564-3},
  \href{http://arxiv.org/abs/1809.04007}{\tt arXiv:1809.04007}.
\bibitem[{Nesvorný and Morbidelli(2012)}]{nesvornymorbidelli12}
\bibinfo{author}{Nesvorný, D.}, \bibinfo{author}{Morbidelli, A.},
  \bibinfo{year}{2012}.
\newblock \bibinfo{title}{{Statistical Study of the Early Solar System's
  Instability with Four, Five, and Six Giant Planets}}.
\newblock \bibinfo{journal}{The Astronomical Journal} \bibinfo{volume}{144},
  \bibinfo{pages}{117}.
\newblock \URLprefix \url{http://stacks.iop.org/1538-3881/144/i=4/a=117}.
\bibitem[{O'Brien et~al.(2006)O'Brien, Morbidelli and Levison}]{obrienetal06}
\bibinfo{author}{O'Brien, D.P.}, \bibinfo{author}{Morbidelli, A.},
  \bibinfo{author}{Levison, H.F.}, \bibinfo{year}{2006}.
\newblock \bibinfo{title}{{Terrestrial planet formation with strong dynamical
  friction}}.
\newblock \bibinfo{journal}{Icarus} \bibinfo{volume}{184},
  \bibinfo{pages}{39--58}.
\newblock \DOIprefix\doi{10.1016/j.icarus.2006.04.005}.
\bibitem[{{Raymond} and {Izidoro}(2017a)}]{raymondizidoro17a}
\bibinfo{author}{{Raymond}, S.N.}, \bibinfo{author}{{Izidoro}, A.},
  \bibinfo{year}{2017}a.
\newblock \bibinfo{title}{{Origin of water in the inner Solar System:
  Planetesimals scattered inward during Jupiter and Saturn's rapid gas
  accretion}}.
\newblock \bibinfo{journal}{\icarus} \bibinfo{volume}{297},
  \bibinfo{pages}{134--148}.
\newblock \DOIprefix\doi{10.1016/j.icarus.2017.06.030},
  \href{http://arxiv.org/abs/1707.01234}{\tt arXiv:1707.01234}.
\bibitem[{{Raymond} and {Izidoro}(2017b)}]{raymondizidoro17b}
\bibinfo{author}{{Raymond}, S.N.}, \bibinfo{author}{{Izidoro}, A.},
  \bibinfo{year}{2017}b.
\newblock \bibinfo{title}{{The empty primordial asteroid belt}}.
\newblock \bibinfo{journal}{Science Advances} \bibinfo{volume}{3},
  \bibinfo{pages}{e1701138}.
\newblock \DOIprefix\doi{10.1126/sciadv.1701138},
  \href{http://arxiv.org/abs/1709.04242}{\tt arXiv:1709.04242}.
\bibitem[{{Raymond} et~al.(2020){Raymond}, {Izidoro} and
  {Morbidelli}}]{raymondetal20}
\bibinfo{author}{{Raymond}, S.N.}, \bibinfo{author}{{Izidoro}, A.},
  \bibinfo{author}{{Morbidelli}, A.}, \bibinfo{year}{2020}.
\newblock \bibinfo{title}{{Solar System Formation in the Context of Extrasolar
  Planets}}, in: \bibinfo{editor}{{Meadows}, V.S.}, \bibinfo{editor}{{Arney},
  G.N.}, \bibinfo{editor}{{Schmidt}, B.E.}, \bibinfo{editor}{{Des Marais},
  D.J.} (Eds.), \bibinfo{booktitle}{Planetary Astrobiology}, p.
  \bibinfo{pages}{287}.
\newblock \DOIprefix\doi{10.2458/azu_uapress_9780816540068}.
\bibitem[{Raymond et~al.(2013)Raymond, Kokubo, Morbidelli, Morishima and
  Walsh}]{raymondetal13}
\bibinfo{author}{Raymond, S.N.}, \bibinfo{author}{Kokubo, E.},
  \bibinfo{author}{Morbidelli, A.}, \bibinfo{author}{Morishima, R.},
  \bibinfo{author}{Walsh, K.J.}, \bibinfo{year}{2013}.
\newblock \bibinfo{title}{{Terrestrial Planet Formation at Home and Abroad}}.
\newblock \bibinfo{journal}{arXiv Prepr. arXiv 1312.1689v3} ,
  \bibinfo{pages}{24}\URLprefix \url{http://arxiv.org/abs/1312.1689},
  \DOIprefix\doi{10.2458/azu{\_}uapress{\_}9780816531240-ch026},
  \href{http://arxiv.org/abs/1312.1689}{\tt arXiv:1312.1689}.
\bibitem[{{Raymond} and {Nesvorn{\'y}}(2022)}]{raymondnesvorny22}
\bibinfo{author}{{Raymond}, S.N.}, \bibinfo{author}{{Nesvorn{\'y}}, D.},
  \bibinfo{year}{2022}.
\newblock \bibinfo{title}{{Origin and Dynamical Evolution of the Asteroid
  Belt}}, in: \bibinfo{booktitle}{Vesta and Ceres. Insights from the Dawn
  Mission for the Origin of the Solar System}, p. \bibinfo{pages}{227}.
\newblock \DOIprefix\doi{10.1017/9781108856324.019}.
\bibitem[{Raymond et~al.(2009)Raymond, O'Brien, Morbidelli and
  Kaib}]{raymondetal09}
\bibinfo{author}{Raymond, S.N.}, \bibinfo{author}{O'Brien, D.P.},
  \bibinfo{author}{Morbidelli, A.}, \bibinfo{author}{Kaib, N.A.},
  \bibinfo{year}{2009}.
\newblock \bibinfo{title}{{Building the terrestrial planets: Constrained
  accretion in the inner Solar System}}.
\newblock \bibinfo{journal}{Icarus} \bibinfo{volume}{203},
  \bibinfo{pages}{644--662}.
\newblock \DOIprefix\doi{10.1016/j.icarus.2009.05.016},
  \href{http://arxiv.org/abs/0905.3750}{\tt arXiv:0905.3750}.
\bibitem[{Raymond et~al.(2004)Raymond, Quinn and Lunine}]{raymondetal04}
\bibinfo{author}{Raymond, S.N.}, \bibinfo{author}{Quinn, T.},
  \bibinfo{author}{Lunine, J.I.}, \bibinfo{year}{2004}.
\newblock \bibinfo{title}{{Making other earths: Dynamical simulations of
  terrestrial planet formation and water delivery}}.
\newblock \bibinfo{journal}{Icarus} \bibinfo{volume}{168},
  \bibinfo{pages}{1--17}.
\newblock \DOIprefix\doi{10.1016/j.icarus.2003.11.019},
  \href{http://arxiv.org/abs/0308159}{\tt arXiv:0308159}.
\bibitem[{Raymond et~al.(2006)Raymond, Quinn and Lunine}]{raymondetal06}
\bibinfo{author}{Raymond, S.N.}, \bibinfo{author}{Quinn, T.},
  \bibinfo{author}{Lunine, J.I.}, \bibinfo{year}{2006}.
\newblock \bibinfo{title}{{High-resolution simulations of the final assembly of
  Earth-like planets I. Terrestrial accretion and dynamics}}.
\newblock \bibinfo{journal}{Icarus} \bibinfo{volume}{183},
  \bibinfo{pages}{265--282}.
\newblock \DOIprefix\doi{10.1016/j.icarus.2006.03.011},
  \href{http://arxiv.org/abs/0510284}{\tt arXiv:0510284}.
\bibitem[{{Roig} et~al.(2021){Roig}, {Nesvorn{\'y}}, {Deienno} and
  {Garcia}}]{roigetal21}
\bibinfo{author}{{Roig}, F.}, \bibinfo{author}{{Nesvorn{\'y}}, D.},
  \bibinfo{author}{{Deienno}, R.}, \bibinfo{author}{{Garcia}, M.J.},
  \bibinfo{year}{2021}.
\newblock \bibinfo{title}{{ISYMBA: a symplectic massive bodies integrator with
  planets interpolation}}.
\newblock \bibinfo{journal}{\mnras} \bibinfo{volume}{508},
  \bibinfo{pages}{4858--4868}.
\newblock \DOIprefix\doi{10.1093/mnras/stab2874},
  \href{http://arxiv.org/abs/2110.00184}{\tt arXiv:2110.00184}.
\bibitem[{de~Sousa~Ribeiro et~al.(2020)de~Sousa~Ribeiro, Morbidelli, Raymond,
  Izidoro, Gomes and {Vieira Neto}}]{ribeiroetal20}
\bibinfo{author}{de~Sousa~Ribeiro, R.}, \bibinfo{author}{Morbidelli, A.},
  \bibinfo{author}{Raymond, S.N.}, \bibinfo{author}{Izidoro, A.},
  \bibinfo{author}{Gomes, R.}, \bibinfo{author}{{Vieira Neto}, E.},
  \bibinfo{year}{2020}.
\newblock \bibinfo{title}{Dynamical evidence for an early giant planet
  instability}.
\newblock \bibinfo{journal}{Icarus} \bibinfo{volume}{339},
  \bibinfo{pages}{113605}.
\newblock \URLprefix
  \url{https://www.sciencedirect.com/science/article/pii/S0019103519301332},
  \DOIprefix\doi{https://doi.org/10.1016/j.icarus.2019.113605}.
\bibitem[{{Vernazza} and {Beck}(2016)}]{vernazzapierre16}
\bibinfo{author}{{Vernazza}, P.}, \bibinfo{author}{{Beck}, P.},
  \bibinfo{year}{2016}.
\newblock \bibinfo{title}{{Composition of Solar System Small Bodies}}.
\newblock \bibinfo{journal}{arXiv e-prints} ,
  \bibinfo{pages}{arXiv:1611.08731}\DOIprefix\doi{10.48550/arXiv.1611.08731},
  \href{http://arxiv.org/abs/1611.08731}{\tt arXiv:1611.08731}.
\bibitem[{{Vokrouhlick{\'y}} et~al.(2016){Vokrouhlick{\'y}}, {Bottke} and
  {Nesvorn{\'y}}}]{vokrouhlickyetal16}
\bibinfo{author}{{Vokrouhlick{\'y}}, D.}, \bibinfo{author}{{Bottke}, W.F.},
  \bibinfo{author}{{Nesvorn{\'y}}, D.}, \bibinfo{year}{2016}.
\newblock \bibinfo{title}{{Capture of Trans-Neptunian Planetesimals in the Main
  Asteroid Belt}}.
\newblock \bibinfo{journal}{\aj} \bibinfo{volume}{152}, \bibinfo{pages}{39}.
\newblock \DOIprefix\doi{10.3847/0004-6256/152/2/39}.
\bibitem[{Walsh et~al.(2011)Walsh, Morbidelli, Raymond, O'Brien and
  Mandell}]{walshetal11}
\bibinfo{author}{Walsh, K.J.}, \bibinfo{author}{Morbidelli, A.},
  \bibinfo{author}{Raymond, S.N.}, \bibinfo{author}{O'Brien, D.P.},
  \bibinfo{author}{Mandell, A.M.}, \bibinfo{year}{2011}.
\newblock \bibinfo{title}{{A low mass for Mars from Jupiter's early gas-driven
  migration.}}
\newblock \bibinfo{journal}{Nature} \bibinfo{volume}{475},
  \bibinfo{pages}{206--209}.
\newblock \URLprefix \url{http://dx.doi.org/10.1038/nature10201},
  \DOIprefix\doi{10.1038/nature10201},
  \href{http://arxiv.org/abs/1201.5177}{\tt arXiv:1201.5177}.
\bibitem[{{Warren}(2011)}]{warren11}
\bibinfo{author}{{Warren}, P.H.}, \bibinfo{year}{2011}.
\newblock \bibinfo{title}{{Stable-isotopic anomalies and the accretionary
  assemblage of the Earth and Mars: A subordinate role for carbonaceous
  chondrites}}.
\newblock \bibinfo{journal}{Earth and Planetary Science Letters}
  \bibinfo{volume}{311}, \bibinfo{pages}{93--100}.
\newblock \DOIprefix\doi{10.1016/j.epsl.2011.08.047}.
\bibitem[{Wetherill(1978)}]{wetherilletal78}
\bibinfo{author}{Wetherill, G.W.}, \bibinfo{year}{1978}.
\newblock \bibinfo{title}{{Accumulation of the terrestrial planets}}, in:
  \bibinfo{editor}{Gehrels, T.} (Ed.), \bibinfo{booktitle}{Protostars planets
  Stud. star Form. Orig. Sol. Syst. (A79-26776 10-90) Tucson}, pp.
  \bibinfo{pages}{565--598}.
\newblock \URLprefix \url{http://adsabs.harvard.edu/abs/1978prpl.conf..565W}.
\bibitem[{{Williams} and {Cieza}(2011)}]{willianscieza11}
\bibinfo{author}{{Williams}, J.P.}, \bibinfo{author}{{Cieza}, L.A.},
  \bibinfo{year}{2011}.
\newblock \bibinfo{title}{{Protoplanetary Disks and Their Evolution}}.
\newblock \bibinfo{journal}{\araa} \bibinfo{volume}{49},
  \bibinfo{pages}{67--117}.
\newblock \DOIprefix\doi{10.1146/annurev-astro-081710-102548},
  \href{http://arxiv.org/abs/1103.0556}{\tt arXiv:1103.0556}.
\bibitem[{Woo et~al.(2023)Woo, Morbidelli, Grimm, Stadel and
  Brasser}]{wooetal23}
\bibinfo{author}{Woo, J.}, \bibinfo{author}{Morbidelli, A.},
  \bibinfo{author}{Grimm, S.}, \bibinfo{author}{Stadel, J.},
  \bibinfo{author}{Brasser, R.}, \bibinfo{year}{2023}.
\newblock \bibinfo{title}{Terrestrial planet formation from a ring}.
\newblock \bibinfo{journal}{Icarus} \bibinfo{volume}{396},
  \bibinfo{pages}{115497}.
\newblock \URLprefix
  \url{https://www.sciencedirect.com/science/article/pii/S001910352300074X},
  \DOIprefix\doi{https://doi.org/10.1016/j.icarus.2023.115497}.
\bibitem[{Yin et~al.(2002)Yin, Jacobsen, Yamashita, Blichert-Toft, T{\'{e}}louk
  and Albar{\`{e}}de}]{yinetal02}
\bibinfo{author}{Yin, Q.}, \bibinfo{author}{Jacobsen, S.B.},
  \bibinfo{author}{Yamashita, K.}, \bibinfo{author}{Blichert-Toft, J.},
  \bibinfo{author}{T{\'{e}}louk, P.}, \bibinfo{author}{Albar{\`{e}}de, F.},
  \bibinfo{year}{2002}.
\newblock \bibinfo{title}{{A short timescale for terrestrial planet formation
  from Hf-W chronometry of meteorites.}}
\newblock \bibinfo{journal}{Nature} \bibinfo{volume}{418},
  \bibinfo{pages}{949--952}.
\newblock \DOIprefix\doi{10.1038/nature00995}.

\end{thebibliography}





\end{document}